\documentclass[12pt]{article}

\usepackage{amssymb,amsthm,amscd, amsbsy, array}
\usepackage{graphicx}
\usepackage{bm}
\usepackage{amsmath,dsfont}

\let\ssection=\section
\renewcommand{\section}{\setcounter{equation}{0}\ssection}

\newcommand{\ba}{{\bf a}}

\newcommand{\balpha}{\boldsymbol{\alpha}}

\newcommand{\alt}{\mathfrak{alt}}

\newcommand{\bb}{{\bf b}}
\newcommand{\bB}{{\bf B}}

\newcommand{\bbeta}{\boldsymbol{\beta}}
\newcommand{\tbeta}{\widetilde{\beta}}

\newcommand{\GammaU}{{}^{U}\!\Gamma}
\newcommand{\gammaU}{{}^{U}\!\gamma}
\newcommand{\cnc}{\mathfrak{cnc}}
\newcommand{\cgal}{\mathfrak{cgal}}
\newcommand{\cmil}{\mathfrak{cmil}}
\newcommand{\tgamma}{\widetilde{\gamma}}

\newcommand{\cD}{{\mathcal{D}}}

\newcommand{\rdiv}{\mathrm{div}}

\newcommand{\bE}{{\bf E}}

\newcommand{\veta}{\boldsymbol{\eta}}

\newcommand{\cF}{{\mathcal{F}}}
\newcommand{\sF}{{\mathcal{F}}}

\newcommand{\bg}{{\bf g}}
\newcommand{\bG}{{\bf G}}
\newcommand{\rg}{\mathrm{g}}

\newcommand{\bgamma}{\boldsymbol{\gamma}}

\newcommand{\gal}{\mathfrak{gal}}

\newcommand{\bj}{{\mathbf{j}}}
\newcommand{\bJ}{{\mathbf{J}}}
\newcommand{\cJ}{{\mathcal{J}}}

\newcommand{\bkappa}{\boldsymbol{\kappa}}

\newcommand{\cM}{\mathcal{M}}

\newcommand{\cO}{{\mathcal{O}}}
\newcommand{\bomega}{{\boldsymbol{\omega}}}

\newcommand{\bp}{{\bf p}}

\newcommand{\bP}{{\bf P}}

\newcommand{\bq}{{\bf q}}

\newcommand{\bbR}{\mathbb{R}}

\newcommand{\bs}{{\bf s}}

\newcommand{\Sch}{\mathrm{Sch}}
\newcommand{\sch}{\mathfrak{sch}}
\newcommand{\Sl}{\mathfrak{sl}}
\newcommand{\SO}{\mathrm{SO}}

\newcommand{\so}{\mathfrak{so}}

\newcommand{\sv}{\mathfrak{sv}}

\newcommand{\dt}{\dot{t}}

\newcommand{\bu}{{\bf u}}

\newcommand{\cU}{{\mathcal{U}}}

\newcommand{\bv}{{\bf v}}

\newcommand{\cV}{{\mathcal{V}}}

\newcommand{\Vect}{\mathrm{Vect}}

\newcommand{\bx}{{\bf x}}

\newcommand{\wX}{{\widetilde{X}}}

\newcommand{\dx}{\dot{x}}
\newcommand{\ddx}{\ddot{x}}

\def\bnabla{{\bm{\nabla}}}
\newcommand{\xx}{x}

\def\Ort{{\rm O}}

\def\valpha{{\bm{\alpha}}}

\def\bu{{\bm{u}}}
\def\bv{{\bm{v}}}
\def\bp{{\bm{p}}}
\def\bP{{\bm{P}}}
\def\bg{{\bm{G}}}
\def\bJ{{\bm{J}}}
\def\bq{{\bm{q}}}

\def\parag{\hfil\break} 

\def\Sch{{{\rm Sch}}}

\def\p{{\partial}}
\def\v0{\mathbf{0}}

\newcommand{\const}{\mathop{\rm const.}\nolimits}
\newcommand{\half }{\frac{1}{2}}

\begin{document}

\title{Non-relativistic conformal symmetries\\ 
and\\  
Newton-Cartan structures 
}

\author{
C. DUVAL\footnote{mailto: duval-at-cpt.univ-mrs.fr}\\
Centre de Physique Th\'eorique, CNRS, 
Luminy, Case 907\\ 
F-13288 Marseille Cedex 9 (France)\footnote{ 
UMR 6207 du CNRS associ\'ee aux 
Universit\'es d'Aix-Marseille I and  II and  Universit\'e du Sud Toulon-Var; Laboratoire 
affili\'e \`a la FRUMAM-FR2291.}
\\[12pt]
P.~A.~HORV\'ATHY\footnote{mailto: horvathy-at-lmpt.univ-tours.fr}\\
Laboratoire de Math\'ematiques et de Physique Th\'eorique\\
Universit\'e de Tours, Parc de Grandmont\\
F-37200 TOURS (France)
}

\date{26 September 2009}

\maketitle
\begin{abstract}
This article provides us with a unifying clas\-sification of the conformal infini\-tesimal sym\-metries of  non-relativistic Newton-Cartan spacetime.
The Lie algebras of non-relativistic conformal transformations
are introduced via the Galilei structure.
They form a family of infinite-dimensional Lie algebras labeled by a rational ``dyna\-mical exponent'', $z$. The Schr\"odinger-Virasoro algebra of Henkel et al. corresponds to $z=2$. Viewed as projective Newton-Cartan symmetries, they yield, for timelike geodesics, the usual Schr\"odinger Lie algebra, for which~$z=2$. For lightlike geodesics, they yield, in turn, the Conformal Galilean Algebra (CGA)  and Lukierski, Stichel and Zakrzewski [alias ``$\alt$" of Henkel], with $z=1$. 
Physical systems realizing these symmetries include, e.g., classical systems of massive, and massless non-relativistic particles, and also hydrodynamics,
 as well as Galilean electromagnetism.
\end{abstract}

\baselineskip=16pt

\texttt{arXiv:0904.0531v5 [hep-th]}.


\newpage
\tableofcontents

\newpage

\section{Introduction}

Non-relativistic conformal symmetries, which are attracting much present interest \cite{LSZGalconf,Henkel03,Henkel06,NRAdS,SZcosmo,Gala,Fouxon,
Gopa,Ali,MaTa,HZ}, are of two types.

Firstly, it has been recognized almost forty years ago \cite{Schr,BPS} that the free Schr\"o\-dinger equation
of a massive particle
has, beyond the obvious Galilean symmetry, two more
``conformal'' symmetries. They are generated by the ``Schr\"odinger'' spacetime vector fields, called \textit{dilation}
\begin{equation}
D=2t\frac{\partial}{\partial t}+x^A\frac{\partial}{\partial x^A}
\label{D}
\end{equation}
and \textit{expansion} (or inversion)
\begin{equation}
K=t^2\frac{\partial}{\partial t}+tx^A\frac{\partial}{\partial x^A}
\label{expan}
\end{equation}
where the dummy index $A$ runs from $1$ to $d$, the dimension of space.

Schr\"odinger dilations and expansions span, with time translations, $H=\partial/{\partial t}$, a Lie algebra isomorphic to $\so(2,1)$. Adding dilations and expansions to the Galilei group yields a two-parameter extension of the latter, dubbed as the (centerless) \textit{Schr\"odinger group}, $\Sch(d)$.\footnote{
The physical realizations of the Schr\"odinger group, in spatial dimension $d\geq3$, admit one more parameter, associated with the \textit{mass}. Adding it yields
the \textit{extended Schr\"odinger group}, which is  the ``non-relativistic conformal'' extension of the one-parameter central extension of  the Galilei group, called the \textit{Bargmann group}. (See, e.g., \cite{DK0,Duv0,DK1,Duv2} for a geometrical account on the Bargmann group.)
In the plane, $d=2$, the Galilei group also has, apart from the previous one, a second, ``exotic'', central extension
widely studied during the last decade \cite{exotic,LSZ,DHexo,HMSP}.
}

Using the word ``conformal''  has been contested \cite{LSZGalconf}, 
hinting at its insufficiently clear 
relation to some conformal structure.
 This  criticism is only half-justified, however. The Schr\"odinger
symmetry has in fact been related to the  Newton-Cartan
structure of non-relativistic spacetime \cite{Duv0,BDP,Duv1,CK}, but this relation has remained rather confidential.  

A different point of view was put forward in Ref. \cite{DBKP},
where it has been shown that non-relativistic theories can be
studied in a ``Kaluza-Klein type'' framework, whereas
the ``non-relativistic conformal'' transformations appear
as those, genuine, conformal transformations of a relativistic
spacetime in one higher dimension, which commute with translations in the ``vertical'' direction. The latter provides us, furthermore, with the central extension required by the mass \cite{BPS,DGH,GPP}.

Secondly, after the pioneering work of Henkel \cite{Hen}, in
\cite{LSZGalconf,Henkel03,Henkel06,NOR-M,Gopa,Ali,MaTa,GL},
attention has been directed to another, less-known and more subtle aspect.
It has been shown, in fact, that a specific group contraction, applied
to the relativistic conformal group $\Ort(d+1,2)$, provides,
for \textit{vanishing mass}, $m=0$, a \textit{second type of conformal extension} of the Galilei group.
 Since group contraction
does not change the number of generators, the new extension, 
called the \textit{Conformal Galilean  Group} \cite{LSZGalconf} has the same
dimension as its relativistic counterpart. Its Lie algebra, the 
\textit{Conformal Galilei Algebra} is spelled as the CGA in the above-mentioned reference.
The CGA is spanned by the vector fields\footnote{The central extensions of the CGA have been discussed in Refs. \cite{LSZGalconf,Henkel06}.}
\begin{equation}
X=\Big(\half \kappa t^2+\lambda t+\varepsilon\Big)\frac{\partial}{\partial t}
+
\Big(
\omega^A_B\,x^B+\lambda{}x^A+\kappa{}t{}x^A-\half \alpha^A{}t^2+\beta^At+\gamma^A\Big)\frac{\partial}{\partial x^A}
\label{CGA} 
\end{equation}
with $\bomega\in\so(d)$, $\balpha,\bbeta,\bgamma\in\bbR^d$, and  $\lambda,\kappa,\varepsilon\in\bbR$. 

The new dilations and expansions, associated with $\lambda$ and $\kappa$ close, with  time translations parametrized by $\varepsilon$, into an~$\so(2,1)$ Lie subalgebra \cite{LSZGalconf,Henkel03,Henkel06,Gopa},
acting \textit{differently} from that of the Schr\"odinger case:
 unlike the ``Schr\"odinger'' one,  (\ref{D}),  the CGA dilation in (\ref{CGA}) dilates space and time at the same rate. Note also the factor~$\half$ in
 the time component of the new expansions.
The vector~$\valpha$ generates, in turn, ``accelerations'' \cite{LSZGalconf}. See also \cite{Hen} for another approach, and \cite{Henkel03} where the CGA was called $\alt(d)$.

The Lie algebra (\ref{CGA}) can be further generalized \cite{Henkel03,Henkel06,Gopa}, in terms of infini\-tesimal ``time redefinition''  and time-dependent translations,
\begin{equation}
X=\xi(t)\frac{\partial}{\partial{}t}+\big(\xi'(t)x^A+\eta^A(t)\big)\frac{\partial}{\partial{}x^A}
\label{infigen}
\end{equation} 
where $\xi(t)$, and~$\veta(t)$ are arbitrary functions of time, $t$. The new expansions and ac\-celera\-tions are plainly recovered choosing
$\xi(t)=\half\kappa\,t^2$ and $\veta(t)=-\half\valpha\,t^2$,
res\-pectively. 
Promoting the infini\-tesimal rotations, $X=\omega^A_B(t)x^B\partial_A$, to be also time-dependent yields an infinite-dimensional conformal extension of the CGA. 

The purpose of the present paper,
a sequel and natural extension of earlier work devoted to Galilean isometries \cite{Duv2}, is to trace-back all these 
``conformal'' symmetries to the \textit{structure of non-relativistic spacetime}. 

Our clue is to
define \textit{non-relativistic conformal transformations} in the framework of Newton-Cartan spacetime \cite{Car,Tra,Kun,DK}, ideally suited to deal with those symmetries in a purely geometric way. In contradistinction to the (pseudo-)Rieman\-nian framework, the degeneracy of the Galilei ``metric'' allows, as we shall see, for infinite-dimensional Lie algebras of conformal Galilei infini\-tesimal transformations, with a wealth of finite-dimensional Lie subalgebras, including the Schr\"odinger Lie algebra and the above-mentioned CGA.

Both the Schr\"odinger
and  Conformal Galilean transformations
turn out to be special cases, related to
our choice of the relative strength of
space and time dilations, characterized by a dynamical exponent \cite{Henkel03,
Henkel06}.

\goodbreak

Our paper is organized as follows.

After reviewing, in Section \ref{NCSection}, the \textit{Newton-Cartan structures} of $(d+1)$-dimen\-sional non-relativistic spacetime, we introduce, in Section \ref{CGalSection}, the notion of \textit{conformal Galilei transformation}. The latter is only concerned with the (singular) ``metric'', $\gamma$, and the ``clock'', represented by a
closed one-form $\theta$. Infinitesimal conformal Galilei trans\-formations form an infinite-dimensional Virasoro-like Lie algebra, denoted by~$\cgal(d)$ in the case of ordinary Galilei spacetime. A geometric definition of the dynamical exponent, $z$, allows us to define the conformal Galilei Lie algebras, $\cgal_z(M,\gamma,\theta)$ of an arbitrary Galilei structure, with prescribed $z$.

Now, Newton-Cartan structures also involve a connection, $\Gamma$, which is \textit{not}
entirely determined by the previous structures. Preserving the geodesic equations
adds, in the generic case, extra conditions, which are explicitly derived in Section \ref{CNCSection}. 

Those help us to reduce the infinite-dimensional conformal Galilei Lie algebra to
that of the Schr\"odinger Lie algebra, $\sch(d)$, for \textit{timelike} geodesics of the flat NC-structure (with dynamical exponent $z=2$). This is reviewed in Section \ref{schSection}.

For \textit{lightlike} geodesics, we get, in turn,  a novel, infinite-dimensional, conformal extension, $\cnc(d)$, of the (centerless) Galilei Lie algebra, which is worked out in Section \ref{NCCLightSection}. This conformal Newton-Cartan Lie algebra admits, indeed, infinite-dimensional Lie subalgebras defined by an arbitrary (rational) dynamical exponent, $z$. Also, the maximal Lie algebra of conformal automorphisms of a \textit{Milne structure}, i.e., a NC-structure with a preferred geodesic and irrotational observer field, shows up as a finite-dimensional Lie algebra, denoted by $\cmil(d)$ in the case of flat spacetime. The CGA (\ref{CGA}) finally appears as a Lie subalgebra of $\cmil(d)$ defined by the dynamical exponent $z=1$. The Lie algebras $\alt_{2/N}(d)$ first defined in \cite{Hen} appear plainly as the Lie algebras of polynomial vector fields of degree $N=1,2,3,\ldots$ in $\cgal_z(d)$ with $z=2/N$. A geometric definition for the latter Lie algebras is still missing, though.

The general theory is illustrated, in Section \ref{dynSys}, on various examples.
Schr\"odinger symmetry is shown to be present
for a Galilean massive particle and in hydro\-dynamics. The massless
non-relativistic particle of Souriau exhibits, as a symmetry, an
infinite-dimensional conformal extension of
the centerless Galilei Lie algebra. At last, the Le Bellac-L\'evy-Leblond theory of (magnetic-like) Galilean electromagnetism carries, apart of
the Schr\"odinger symmetry, also the CGA.

\goodbreak

\section{Newton-Cartan structures}\label{NCSection}

\subsection{Galilei structures and Newton-Cartan connections}

Let us recall that a Newton-Cartan (NC) spacetime structure, $(M,\gamma,\theta,\Gamma)$, consists of a smooth, connected, $(d+1)$-dimensional manifold $M$, a twice-contravariant sym\-metric tensor field
$\gamma=\gamma^{ab}\,\partial_a\otimes\partial_b$
(where $a,b=0,1,\ldots,d$) of signature $(0,+,\ldots,+)$ whose kernel is spanned by the 
one-form
$\theta=\theta_adx^a$.
Also $\Gamma$ is a Galilei connection, i.e., a symmetric linear connection compatible with $\gamma$ and $\theta$
\cite{Car,Tra,Kun,DK,Kun1,Lin}. 

\goodbreak

Now, in contra\-distinction to the relativistic framework, such a connection is \textit{not}
uniquely determined by the Galilei spacetime structure $(M,\gamma,\theta)$.
Therefore, in order to reduce the ambiguity, one usually introduces NC-connection as Galilei connections subject to the nontrivial
sym\-metry of the curvature: $R_{a\ c}^{\ b\ d}=R_{c\ a}^{\ d\ b}$ 
(where $R_{a\ c}^{\ b\ d}\equiv\gamma^{bk}\,R_{akc}^{\ \ \ d}$); the latter may be
thought of as part of the covariant Newtonian gravitational field equations~\cite{Tra,Kun,DK0,DK}. 

\goodbreak

Under mild geometric conditions, the quotient $T=M/\ker(\theta)$ is a well-behaved one-dimensional manifold,
interpreted as the time axis endowed with the closed one-form $\theta$, interpreted as the Galilei \textit{clock}. The tensor field $\gamma$ then defines a Riemannian metric on each of the (spacelike) fibers of the projection $M\to{}T$.
 
\goodbreak

The standard example of a NC-structure is given by
$M\subset\bbR\times\bbR^d$ together with
$\gamma=\delta^{AB}\partial_A\otimes\partial_B$ (where $A,B=1,\ldots,d$), and $\theta=dx^0$; the nonzero components of the connection,
$\Gamma_{00}^{A}=\partial_AV$,
host the Newtonian scalar potential, $V$. The above coordinate system $(x^0,\ldots,x^d)$ will be called Galilean. 

\goodbreak

The {\it flat} NC-structure corresponds to the subcase where $M=\bbR\times\bbR^d$, and
\begin{equation}
\gamma^{ab}=\delta^a_A\delta^b_B\,\delta^{AB},
\qquad
\theta_a=\delta^0_a,
\qquad
\Gamma^c_{ab}=0
\label{flatNC}
\end{equation}
for all $a,b,c=0,\ldots,d$. Such a coordinate system will be called (NC-)inertial.

\goodbreak

Since we will be dealing with ``conformal'' Galilean spacetime transformations that preserve the directions of the Galilei structure, we must bear in mind that the transformation law of the NC-connection, $\Gamma$, will have to be specified independently of that of the Galilei ``metric'' $(\gamma,\theta)$, which is clearly due to the fact that there are extra degrees of freedom associated with NC-connections. Let us, hence, describe the precise geometric content of NC-connections.

\goodbreak 

It has been shown \cite{Kun} that NC-connections can be decomposed according to\footnote{Round brackets denote symmetrization, and square ones will denote skew-symmetrization.}
\begin{equation}
\Gamma_{ab}^{c}
=
\GammaU_{ab}^{c}+\theta_{(a}F_{b)k}\gamma^{kc}
\label{Gamma}
\end{equation}
where \cite{Tru}
\begin{equation}
{\GammaU}_{ab}^{c}
=
\gamma^{ck}\Big(
\partial_{(a}\gammaU_{b)k} - \half \partial_{k}\gammaU_{ab}
\Big)
+
\partial_{(a}\theta_{b)}\,U^c 
\label{GammaU}
\end{equation}
is the unique NC-connection for which the unit space\-time vector field $U$
(i.e., such that $\theta_aU^a=1$) is geodesic and curlfree, $F$ being an otherwise 
arbitrary closed two-form. 
Here $\gammaU$ is the symmetric, twice-covariant, tensor field uniquely determined by
$\gammaU_{ak}\gamma^{kb}=\delta_a^b -U^b\theta_a$ and
$\gammaU_{ak}U^k=0$. From a mechanical standpoint, the above two-form, $F$, of $M$ encodes Coriolis-like accelerations relatively to the observer~$U$. 

\goodbreak

For example, if $M\subset\bbR\times\bbR^3$, the constant, future-pointing, vector field $U=\partial_0$ will represent the four-velocity of an observer. Now, $F$ being closed, one has, locally, $F=dA$ for some one-form $A$, e.g., $A=-V(t,\bx)dt+\omega(t)_{BC}\,x^Bdx^C$, where $V(t,\bx)$ is the Newtonian (plus centrifugal) potential, and $\bomega(t)\in\so(3)$ the time-dependent angular velocity of the observer relatively to the Galilei frame associated with the coordinates $t=x^0$, and $\bx=(x^1,x^2,x^3)$. Anticipating the equations of free fall, we check that the equations of NC-geodesics (\ref{EqGeod}) --- with the choice of time, $t$, as an affine parameter --- yield, with the help of (\ref{Gamma}), the familiar equations 
\begin{equation}
\ddot{t}=0,
\qquad
\ddot\bx=-\bnabla{V}+\dot\bomega\times\bx+2\bomega\times\dot\bx
\end{equation} 
governing the motion of a massive particle in a rotating Galilei coordinate system.\footnote{The non-trivial components of the NC-connection (\ref{Gamma}) read, in this case, $\Gamma^A_{00}=\partial_AV-\dot\omega^A_Bx^B$, and $\Gamma^A_{B0}=-\omega^A_B$, for all $A,B=1,2,3$.}

\subsection{NC-gauge transformations, and NC-Milne structures}\label{NCGaugeSection}

\subsubsection{Gauge transformations}

We have seen that, in view of~(\ref{Gamma}), we can usefully parametrize NC-connections, $\Gamma$, by the previously introduced pairs~$(U,F)$ which are, themselves, not entirely fixed by the NC-connection. (This arbitrariness in the expression of the NC-connection can be traced-back to the degeneracy of the Galilei structure ; this does not occur in the pseudo-Riemannian case where the Levi-Civita connection is uniquely determined by the metric.)

Let us mention \cite{Kun,DK} that for a given, \textit{fixed}, Galilei structure $(\gamma,\theta)$, the pair~$(U',F')$ defines the same NC-connection, $\Gamma$, as $(U,F)$ does iff both are \textit{gauge}-related by a so-called \textit{Milne boosts} \cite{CK,Duv2}
\begin{equation}
U'=U+\gamma(\Psi),
\qquad
F'=F+d\Phi
\label{Gauge}
\end{equation}
where $\Psi=\Psi_adx^a$ is an arbitrary one-form of $M$, which may be interpreted as a boost,\footnote{Two observers $U$, and $U'$ are related by a boost, i.e., an acceleration $A=U'-U$ which is necessarily spacelike, $\theta(A)=0$, hence of the form $A=\gamma(\Psi)$, as specified in (\ref{Gauge}).
}
and $\Phi=\Phi_adx^a$ is such that
\begin{equation}
\Phi_a=\Psi_a-\Big(\Psi_bU^b+\half \gamma^{bc}\Psi_b\Psi_c\Big)\theta_a.
\label{Phi}
\end{equation}
The infini\-tesimal versions of the preceding gauge transformations read, accordingly,
\begin{equation}
\delta{U}=\gamma(\psi),
\qquad
\delta{F}=d\phi
\label{gauge}
\end{equation}
where $\psi$ is an arbitrary one-form  of $M$ (an infinitesimal boost), and\begin{equation}
\phi=\gammaU(\delta{U}).
\label{phi}
\end{equation}
One readily checks that, indeed, $\delta\Gamma=0$.

\subsubsection{NC-Milne structure}

In fact, given a NC-connection, $\Gamma$, and an arbitrary observer, $U$, one uniquely determines the (closed) ``Coriolis'' two-form, $F$, via the fundamental relation \cite{Kun,DK}
\begin{equation}
F_{ab}=-2\,\gammaU_{c[a}\nabla_{b]}U^c
\label{F}
\end{equation}
where $\nabla$ stands for the covariant derivative associated with the NC-connection, $\Gamma$.
This implies that the \textit{geodesic acceleration}, $\dot{U}^a=U^b\nabla_bU^a$, of the observer $U$ reads
\begin{equation}
\dot{U}^a=-F^a_bU^b
\label{dotU}
\end{equation}
while its \textit{curl} is of the form
\begin{equation}
2\nabla^{[a}U^{b]}=F^{ab}
\label{curlU}
\end{equation}
where coordinate indices have lifted using $\gamma$, e.g., $F^a_b=\gamma^{ac}F_{cb}$. An inertial and non-rotating observer, $U$, will therefore be characterized by $F=0$. Whenever such an observer exists, it will be called an \textit{ether}, in the spirit of \cite{CK}.

We call \textit{NC-Milne structure} a NC-structure admitting an observer $U$ such that
\begin{equation}
F_{ab}=0
\label{F=0}
\end{equation}
for all $a,b=0,\ldots,d$. We will denote this special NC-structure by $(M,\gamma,\theta,\GammaU)$; see~(\ref{Gamma}). 

\goodbreak

\section{Conformal Galilei transformations, Schr\"o\-dinger-Virasoro Lie algebra}
\label{CGalSection}

\subsection{The Lie algebra, $\cgal$, of conformal Galilei transformations}

In close relationship to the Lorentzian framework, we call \textit{conformal Galilei} transformation of $(M,\gamma,\theta)$ any diffeomorphism of $M$ that preserves the direction of~$\gamma$. Owing to the fundamental constraint $\gamma^{ab}\theta_b=0$, it follows that conformal Galilei transformation automatically preserve the direction of the time one-form~$\theta$.

In terms of infinitesimal transformations, a \textit{conformal Galilei} vector field of $(M,\gamma,\theta)$ is a vector field, $X$, of $M$ that Lie-transports the direction of $\gamma$; we will thus define $X\in\cgal(M,\gamma,\theta)$ iff
\begin{equation}
L_X\gamma=f\gamma
\qquad
\hbox{hence}
\qquad
L_X\theta=g\,\theta
\label{confgal} 
\end{equation}
for some smooth functions $f,g$ of $M$, depending on $X$. Then, $\cgal(M,\gamma,\theta)$ becomes a Lie algebra whose bracket is the Lie bracket of vector fields.

The one-form $\theta$ being parallel-transported by the NC-connection, one has neces\-sarily $d\theta=0$; this yields $dg\wedge\theta=0$, implying that $g$ is (the pull-back of) a smooth function on~$T$, i.e., that $g(t)$ depends arbitrarily on time $t=x^0$, which locally parametrizes the time axis. We thus have $dg=g'(t)\theta$.

Let us work out the expression of the generators of the \textit{conformal Galilei Lie algebra}, $\cgal(d)=\cgal(\bbR\times\bbR^d,\gamma,\theta)$, of the flat NC-structure (\ref{flatNC}). Those are the vector fields, $X=X^0\partial_0+X^A\partial_A$, solutions of (\ref{confgal}), namely such that\footnote{Let us recall the general expressions of the Lie derivatives of $\gamma$ and $\theta$ along the vector field $X=X^a\partial_a$ of $M$, namely $L_X\gamma^{ab}=X^c\partial_c\gamma^{ab}-2\partial_cX^{(a}\gamma^{b)c}$, and $L_X\theta_a=\partial_a(\theta_bX^b)$.}
\begin{eqnarray}
\label{LXgammaAB}
\partial_AX_B+\partial_BX_A&=&-f\,\delta_{AB}\\
\partial_AX^0&=&0\\
\label{LXtheta0}
\partial_0X^0&=&g
\end{eqnarray}
for all $A,B=1,\ldots,d$. (We have put $X_A=\delta_{AB}X^B$.)


We readily find that $X\in\cgal(d)$ iff\footnote{We will assume $d>1$.}
\begin{equation}
X=\xi(t)
\frac{\partial}{\partial t}
+
\Big(\omega^A_B(t)x^B+\eta^A(t)+\kappa^A(t)x_Bx^B-2x^A\kappa_B(t)x^B+\chi(t)x^A\Big)\frac{\partial}{\partial x^A}
\label{cgal} 
\end{equation}
where $\bomega(t)\in\so(d)$, $\veta(t)$, $\bkappa(t)$, $\chi(t)$, and $\xi(t)$ are arbitrary functions of time, $t$; those are clearly interpreted as time-dependent infinitesimal rotations, space translations, expansions (or inversions), space dilations, and time reparametrizations.

We note, \textsl{en passant}, that the $\cgal(d)$-generators (\ref{cgal}) project as vector fields of the time axis; therefore, there exists a canonical Lie algebra homomorphism: $\cgal(d)\to\Vect(\bbR)$ given by $X\mapsto{}\xi(t)
\partial_t$, onto the Lie algebra of vector fields of~$T\cong\bbR$, i.e., the (centerless) \textit{Virasoro Lie algebra}.

\subsection{Conformal Galilei transformations, $\cgal_z$, with dynamical ex\-ponent~$z$}\label{cgalzSection}

One can, at this stage, try and seek non-relativistic avatars of general relativistic infinitesimal conformal transformations. Given a Lorentzian (ore, more generally, a pseudo-Riemannian) manifold $(M,\rg)$, the latter Lie algebra is generated by the vector fields, $X$, of~$M$ such that
\begin{equation}
L_X(\rg^{-1}\otimes\rg)=0
\label{conf} 
\end{equation}
where $\rg^{-1}$ denotes the inverse of the metric $\rg:TM\to{}T^*M$.

\goodbreak

It has been shown \cite{Kun1} that one can expand a Lorentz metric in terms of the small parameter $1/c^2$, where $c$ stands for the speed of light, as  $\rg=c^2\theta\otimes\theta-\gammaU+\cO(c^{-2})$, and $\rg^{-1}=-\gamma+c^{-2}U\otimes{}U+\cO(c^{-4})$, with the same notation as before. Then, a non-relativistic limit of Equation (\ref{conf}) would be $L_X\lim_{c\to\infty}(c^{-2}\,\rg^{-1}\otimes\rg)=0$, viz.,
\begin{equation}
L_X(\gamma\otimes\theta\otimes\theta)=0.
\label{q=2} 
\end{equation}
This is merely one of the possibilities at hand in our formalism. In fact, having at our disposal a Galilei structure on $M$, we will introduce, instead of (\ref{q=2}), a more flexible condition. Indeed, owing to the degeneracy of the Galilei ``metric''~$(\gamma,\theta)$, we will deal with the following condition, namely,
\begin{equation}
L_X(\gamma^{\otimes{}m}\otimes\theta^{\otimes{}n})=0
\label{galconfMN} 
\end{equation}
for some $m=1,2,3,\ldots$, and $n=0,1,2,\ldots$, to be further imposed on the vector fields $X\in\cgal(M,\gamma,\theta)$. This is equivalent to Equation (\ref{confgal}) together with the extra condition
\begin{equation}
f+q\,g=0\qquad
\hbox{where}
\qquad
q=\frac{n}{m}.
\label{f+qg=0} 
\end{equation}
Indeed, $L_X(\gamma^{\otimes{}m}\otimes\theta^{\otimes{}n})=0$ implies $L_X\gamma=f\gamma$ and $L_X\theta=g\,\theta$ for some functions~$f$ and $g$ of $M$ such that $mf+ng=0$. 
 Equation (\ref{q=2}) plainly corresponds to the special case $m=1$, $n=2$.
 
From now on, we will call \textit{dynamical exponent} the quantity
\begin{equation}
z=\frac{2}{q}
\label{z} 
\end{equation}
where $q$ is as in (\ref{f+qg=0}). This quantity will be shown to match the ordinary notion of dynamical exponent; see, e.g., \cite{Hen,Henkel03}.

We will, hence, introduce the Galilean avatars, $\cgal_z(M,\gamma,\theta)$, of the Lie algebra~$\so(d+1,2)$ of conformal vector fields of a pseudo-Riemannian structure of signature $(d,1)$ as the Lie algebras spanned by the vector fields $X$ of $M$ satisfying~(\ref{confgal}), and (\ref{galconfMN}) --- or (\ref{f+qg=0}) for some rational number $z$. We will call $\cgal_z(M,\gamma,\theta)$ the \textit{conformal Galilei Lie algebra with dynamical exponent} $z$ (see (\ref{z})).

The Lie algebra 
\begin{equation}
\sv(M,\gamma,\theta)=\cgal_2(M,\gamma,\theta)
\label{sv}
\end{equation}
is the obvious generalization to Galilei spacetimes of the \textit{Schr\"odinger-Virasoro} Lie algebra $\sv(d)=\sv(\bbR\times\bbR^d,\gamma,\theta)$ introduced in \cite{Hen} (see also \cite{Henkel03}) from a different viewpoint in the case of a flat NC-structure.  The representations of the Schr\"odinger-Virasoro group and of its Lie algebra, $\sv(d)$, as well as the deformations of the latter have been thoroughly studied and investigated in \cite{RU}. 

An easy calculation using (\ref{LXgammaAB}), (\ref{LXtheta0}), the new constraint (\ref{f+qg=0}), and (\ref{z}) shows that $X\in\cgal_z(d)$ iff
\begin{equation}
X=\xi(t)\frac{\partial}{\partial t}
+
\Big(\omega^A_B(t)x^B+\frac{1}{z}\xi'(t)x^A+\eta^A(t)\Big)\frac{\partial}{\partial x^A}
\label{cgalz} 
\end{equation}
where $\bomega(t)\in\so(d)$, $\veta(t)$, and $\xi(t)$ depend arbitrarily on time, $t$. 
Equation (\ref{cgalz}) generalizes (\ref{infigen}) from $z=1$ to any $z$.

The Lie algebra $\cgal_\infty(M,\gamma,\theta)$ corresponding to the case $q=0$ is interesting (see below, Section \ref{Gal30sSection})). We have, indeed, $X\in\cgal_\infty(M,\gamma,\theta)$ iff 
\begin{equation}
L_X\gamma=0.
\label{LXgamma=0} 
\end{equation}
In the case of a flat NC-structure, $\cgal_\infty(d)$ is spanned by the vector fields
\begin{equation}
X=\xi(t)
\frac{\partial}{\partial t}
+
\Big(\omega^A_B(t)x^B+\eta^A(t)\Big)\frac{\partial}{\partial x^A}
\label{zinfinity} 
\end{equation}
where, again, $\bomega(t)\in\so(d)$, $\veta(t)$, and $\xi(t)$ depend arbitrarily on time, $t$.

\section{Conformal Newton-Cartan transformations}\label{CNCSection}

As previously emphasized, NC-connections are quite independent geometric objects; they, hence, deserve a special treatment. The idea pervading earlier work \cite{Duv0,Duv1,Duv2} on non-relativistic symmetries is that specifying explicitly the transformation law of the NC-connection is mandatory in a number of cases, e.g., those relevant to geometric mechanics and non-relativistic physical theories.

We will, henceforth, focus attention on the notion of Newtonian geodesics; more particularly, we will insist that the above-mentioned Galilean conformal trans\-formations should, in addition, permute the NC-geodesics.

\goodbreak

The geodesics of a NC-structure $(M,\gamma,\theta,\Gamma)$ are plainly geodesics of $(M,\Gamma)$, i.e., the solutions of the differential equations
\begin{equation}
\ddx^c+\Gamma^c_{ab}\,\dx^a\dx^b=\mu\,\dx^c
\label{EqGeod}
\end{equation}
for all $c=0,\ldots,d$, where $\mu$ is some smooth (fiberwise linear) function of $TM$; here, we have put $\dx^a=dx^a/d\tau$, where $\tau$ is an otherwise arbitrary curve-parameter. 

Let us remind that Equation~(\ref{EqGeod}) models \textit{free fall} in NC theory \cite{Car,Tra,Kun}, just as it does in general relativity.
By putting~$\dt=\theta_a\dx^a$, we characterize\footnote{The condition $\dt=0$ is clearly a first-integral of Equation (\ref{EqGeod}). Lightlike --- or \textit{null} --- geodesics are, hence, spacelike;
the origin of the terminology will be explained later, in Section \ref{dynSys}.}
\begin{eqnarray}
\hbox{\textit{timelike} geodesics by:}
&\dt\neq0
\label{timelikegeod}
\\[6pt]
\hbox{\textit{lightlike} geodesics by:}
&\dt=0.
\label{spacelikegeod}
\end{eqnarray}

\goodbreak

Spacetime transformations which permute the geodesics of~$(M,\Gamma)$, i.e., preserve the form of the geodesic equation (\ref{EqGeod}), are \textit{projective} transformations; they form the projective group of the affine structure. Infinitesimal projective trans\-formations generate a Lie algebra which, hence, consists of vector fields, $X$, of $M$ satisfying
\begin{equation}
L_X\Gamma^c_{ab}=\delta^c_a\varphi^{}_b+\delta^c_b\varphi^{}_a
\label{TransfProj}
\end{equation}
for a certain one-form $\varphi=\varphi_adx^a$ of $M$ depending on $X$.

\goodbreak

\subsection{The Schr\"odinger Lie algebra}
\label{schSection}

Let us first cope with generic, \textit{timelike}, geodesics of $(M,\Gamma)$ defined by $\dt\neq0$, cf.~Equation (\ref{timelikegeod}), and representing the worldlines of \textit{massive non-relativistic test particles}. From now on, we choose to enforce preservation of their equations~(\ref{EqGeod}), in addition to that, (\ref{confgal}), of the direction of the Galilei structure~$(\gamma,\theta)$. 

\subsubsection{The expanded Schr\"odinger Lie algebra, $\widetilde{\sch}$, of projective Galilei conformal transformations}

The Lie-transport (\ref{TransfProj}) of the NC-connection, compatible with the conformal re\-scalings (\ref{confgal}) of the Galilei structure $(\gamma,\theta)$, must preserve the first constraint~$\nabla\theta=0$, i.e., $L_X\nabla_a\theta_b=\nabla_aL_X\theta_b-\theta_cL_X\Gamma^c_{ab}=0$; this yields $g'\theta_a\theta_b-2\theta_{(a}\varphi_{b)}=0$, or $\varphi_a=\half{}g'\theta_a$. 
The infinitesimal projective transformations to consider are thus given by
\begin{equation}
L_X\Gamma^c_{ab}
=
g'\delta^c_{(a}\theta^{}_{b)}.
\label{deltaGammaSch}
\end{equation}
Likewise, preservation of the second constraint, viz., $\nabla\gamma=0$, necessarily implies $L_X\nabla_c\gamma^{ab}=\nabla_c\,L_X\gamma^{ab}+2L_X\Gamma^{(a}_{ck}\gamma^{b)k}=0$; we thus find $(\partial_cf+g'\theta_c)\gamma^{ab}=0$, and $f$ is therefore a function of $T$ such that
\begin{equation}
f'+g'=0.
\label{fp+gp=0} 
\end{equation}

\goodbreak

We will, hence, define a new Lie algebra, denoted $\widetilde{\sch}(M,\gamma,\theta,\Gamma)$, as the Lie algebra of those vector fields that are infinitesimal (i) conformal Galilei transformations of $(M,\gamma,\theta)$, and (ii)~projective transformations of $(M,\Gamma)$. We call $\widetilde{\sch}(M,\gamma,\theta,\Gamma)$ the \textit{expanded Schr\"odinger Lie algebra}, which is therefore spanned by the vector fields, $X$, of~$M$ such that \cite{Duv0,BDP,Duv1}
\begin{equation}
L_X\gamma^{ab}=f\gamma^{ab},
\qquad
L_X\theta_a=g\,\theta_a
\qquad
\&
\qquad
L_X\Gamma^c_{ab}
=
g'\delta^c_{(a}\theta^{}_{b)}
\label{sch} 
\end{equation}
for all $a,b,c=0,1,\ldots,d$, and subject to Condition (\ref{fp+gp=0}).

Let us now work out the form of the Schr\"odinger Lie algebra in the flat case. We will thus determine the generators of the Lie algebra $\widetilde{\sch}(d)=\widetilde{\sch}(\bbR\times\bbR^d,\gamma,\theta,\Gamma)$ in the special case (\ref{flatNC}).
The system (\ref{sch}) to solve for $X=X^0\partial_0+X^A\partial_A$ reads\footnote{Let us recall that the Lie derivative of a linear connection, $\Gamma$, along the vector field, $X$, is given by $L_X\Gamma^c_{ab}=\partial_a\partial_bX^c$ in the \textit{flat} case, and in a coordinate system where $\Gamma^c_{ab}=0$.}
\begin{eqnarray}
\label{1}
\partial_AX_B+\partial_BX_A&=&-f\delta_{AB}\\
\label{2}
\partial_AX^0&=&0\\
\label{3}
\partial_0X^0&=&g\\
\label{6}
\partial_0\partial_0X^A&=&0\\
\label{7}
\partial_0\partial_BX^A&=&\half {}g'\delta^A_B\\
\label{9}
\partial_A\partial_BX^C&=&0
\end{eqnarray}
for all $A,B,C=1,\ldots,d$. 

\goodbreak

We deduce, from (\ref{9}) that 
$X^A=M^A_B(t)x^B+\eta^A(t)$, and, using (\ref{1}), we find 
$M^A_B(t)=\omega^A_B-\half{}f(t)\delta^A_B$, 
where the $\omega_{AB}=-\omega_{BA}$ are independent of $t$. Then~(\ref{6}) leaves us with 
$f''(t)=0$, and $(\eta^A)''(t)=0$, i.e., with $f(t)=-2(\kappa t+\lambda)$, and $\eta^A(t)=\beta^A t+\gamma^A$, where $\kappa,\lambda,\beta^A$, and $\gamma^A$ are constant coefficients. At last, using Equations~(\ref{fp+gp=0}) and (\ref{3}), we conclude that $X^0=\kappa t^2+\mu{}t+\varepsilon$, with $\mu,\varepsilon$ new constants of integration. 
  
We can therefore affirm that $X\in\widetilde{\sch}(d)$ iff
\begin{equation}
X
=
\left(\kappa{}t^2+\mu{}t+\varepsilon\right)\frac{\partial}{\partial t}
+
\left(\omega^A_B\,x^B+\kappa{}t{}x^A+\lambda{}x^A+\beta^A{}t+\gamma^A\right)\frac{\partial}{\partial x^A}
\label{schd} 
\end{equation}
where $\bomega\in\so(d)$, $\bbeta,\bgamma\in\bbR^d$, and $\kappa,\mu,\lambda,\varepsilon\in\bbR$ are respectively infinitesimal rotations, boosts, spatial translations, inversions, time dilations, space dilations, and time translations. We observe in (\ref{schd}) that time is dilated \textit{independently} of space~\cite{Duv0,Duv1}. The expanded Schr\"odinger Lie algebra, $\widetilde{\sch}(d)$, is a finite-dimensional Lie subalgebra of $\cgal(d)$.


Let us recall that the \textit{Galilei Lie algebra} $\gal(M,\gamma,\theta,\Gamma)\subset\cgal(M,\gamma,\theta,\Gamma)$ of a NC-structure is plainly defined as its Lie algebra of infinitesimal automorphisms. Thus, $X\in\gal(M,\gamma,\theta,\Gamma)$ iff \cite{Tra,Duv2}
\begin{equation}
L_X\gamma^{ab}=0,
\qquad
L_X\theta_a=0
\qquad
\&
\qquad
L_X\Gamma^c_{ab}
=
0
\label{gal} 
\end{equation}
for all $a,b,c=0,1,\ldots,d$, i.e., if (\ref{sch}) holds with $f=0$, and $g=0$. In the flat case, $\gal(d)=\gal(\bbR\times\bbR^d,\gamma,\theta,\Gamma)$ is clearly spanned by the vector fields (\ref{schd}) with $\kappa=0$, and $\lambda=\mu=0$.

\subsubsection{The Schr\"odinger Lie algebras, $\sch_z$, with dynamical exponent $z$}\label{schzSection}

Just as in Section \ref{cgalzSection}, we define \textit{Schr\"odinger Lie algebra with dynamical exponent $z$} as the Lie subalgebra $\sch_z(M,\gamma,\theta,\Gamma)\subset\widetilde{\sch}(M,\gamma,\theta,\Gamma)$ defined by the supplementary condition (\ref{f+qg=0}), i.e.,
\begin{equation}
f+\frac{2}{z}g=0
\label{f+2g/z=0}
\end{equation}
where $z$ is given by (\ref{z}). 
This entails, via Equation (\ref{fp+gp=0}), that 
\begin{equation}
\Big(\frac{2}{z}-1\Big)g'(t)=0.
\label{zfixingcond}
\end{equation}

\goodbreak

$\bullet$  We, hence, find
\begin{equation}
z=2
\label{z=2}
\end{equation}
since $g'\neq0$, \textit{generically}.
For the flat NC-structure, see (\ref{schd}), this implies that time is dilated \textit{twice} as much as space \cite{Schr}, a specific property of the (centerless) \textit{Schr\"odinger Lie algebra} 
\begin{equation}
\sch(d)=\sch_2(d)
\label{schDef}
\end{equation}
for which \begin{equation}
\mu=2\lambda.
\label{mu=2lambda}
\end{equation}

\goodbreak

We therefore contend that $X\in\sch_2(d)$ iff
\begin{equation}
X
=
\left(\kappa{}t^2+2\lambda{}t+\varepsilon\right)\frac{\partial}{\partial t}
+
\left(\omega^A_B\,x^B+\kappa{}t{}x^A+\lambda{}x^A+\beta^A{}t+\gamma^A\right)\frac{\partial}{\partial x^A}
\label{sch2d} 
\end{equation}
where $\bomega\in\so(d)$, $\bbeta,\bgamma\in\bbR^d$, and $\kappa,\lambda,\varepsilon\in\bbR$.
The Schr\"odinger \textit{dynamical exponent} is $z=2$; see, e.g., \cite{Hen}.  

The Lie algebra $\sch(d)$ admits the faithful $(d+2)$-dimensional representation $X\mapsto{}Z$ where
\begin{equation}
Z=
\left(
\begin{array}{rrr}
\bomega&\bbeta&\bgamma\\[6pt]
0&\lambda&\varepsilon\\[6pt]
0&-\kappa&-\lambda
\end{array}
\right)
\label{sch2dRep} 
\end{equation}
with the same notation as above. 

\goodbreak

We therefore have the Levi decomposition
\begin{equation}
\sch(d)\cong(\so(d)\times\Sl(2,\bbR))\ltimes(\bbR^d\times\bbR^d).
\label{sch2dLeviMalcev} 
\end{equation}

The Schr\"odinger Lie algebra is, indeed, a \textit{finite-dimensional} Lie sub\-algebra of the Schr\"odinger-Virasoro Lie algebra (\ref{sv}), viz., 
\begin{equation}
\sch_2(M,\gamma,\theta,\Gamma)\subset\sv(M,\gamma,\theta).
\label{schsubsetschVir} 
\end{equation}

$\bullet$ Returning to Equation (\ref{zfixingcond})
 we get, in the special case $g'=0$, a family of Lie subalgebras algebras $\sch_z(M,\gamma,\theta,\Gamma)\subset\widetilde{\sch}(M,\gamma,\theta,\Gamma)$ parametrized by a (rational) dynamical exponent, $z$. In the flat case, $\sch_{z\neq2}(d)$ is spanned by the vector fields~(\ref{schd}) with $\kappa=0$, and $\mu=z\lambda$.

In the limit $z\to\infty$, where $f=g'=0$ in view of (\ref{fp+gp=0}), we obtain the Lie algebra $\sch_\infty(M,\gamma,\theta,\Gamma)$. For flat NC-spacetime, $\sch_\infty(d)$ is generated by the vector fields~(\ref{schd}) with $\kappa=\lambda=0$.

In both cases we get the Lie algebra of vector fields of the form
\begin{equation}
X
=\left(
\mu{}t+\varepsilon\right)\frac{\partial}{\partial t}
+
\left(\omega^A_B\,x^B+\frac{\mu}{z}\,x^A+\beta^A{}t+\gamma^A\right)\frac{\partial}{\partial x^A}
\label{genzsch} 
\end{equation}
with the same notation as above.

\subsection{Transformation law of NC-connections under conformal Galilei rescalings}

The rest of the section will be devoted to the specialization of projective transformations to the specific case of lightlike (\ref{spacelikegeod}) NC-geodesics.

Let us now work out the general form of the variation, $\delta\Gamma$, of a NC-connection, $\Gamma$, under infinitesimal conformal rescalings of the Galilei structure $(\gamma,\theta)$ of $M$, namely
\begin{equation}
\delta\gamma=f\gamma
\qquad
\mbox{hence}
\qquad
\delta\theta=g\,\theta
\label{deltagammadeltatheta}
\end{equation}
where $f$ is an arbitrary function of $M$, and $g$ an arbitrary function of $T$ (compare Equation (\ref{confgal})). We will furthermore put, in full generality,
\begin{equation}
\delta{U}=-gU+\gamma(\psi)
\label{deltaU}
\end{equation}
in order to comply with the constraint $\theta_aU^a=1$, where $\psi$ is an arbitrary one-form of $M$ interpreted as an infinitesimal Milne boost (cf. (\ref{gauge})).

\goodbreak

Starting from (\ref{Gamma}), we get
$
\delta\Gamma^c_{ab}=(\delta\GammaU)^c_{ab}+\delta\theta_{(a}F_{b)k}\gamma^{ck}+\theta_{(a}\delta{F}_{b)k}\gamma^{ck}+\theta_{(a}F_{b)k}\delta\gamma^{ck}
$. Then, using (\ref{deltagammadeltatheta}) and (\ref{deltaU}) applied to the expression (\ref{GammaU}) of the NC-connection~$\GammaU$, we find
$
\delta(\GammaU)^c_{ab}
=
-\delta^c_{(a}\partial^{}_{b)}f
+U^c\theta_{(a}\partial_{b)}(f+g)
+\half \left(\gamma^{ck}\partial_kf\right)\gammaU_{ab}
-\theta_{(a}d\phi_{b)k}\gamma^{ck}
$,
where $\phi=\gammaU(\delta{}U)$ is the one-form (\ref{phi}) associated with the ``Milne'' variation (\ref{deltaU}) of the observer $U$.
Then, with the help of Equation (\ref{deltagammadeltatheta}), and of general result $\delta{F}=d\phi$, see (\ref{gauge}), we can finally claim that
\begin{equation}
\delta\Gamma^c_{ab}
=
-\delta^c_{(a}\partial^{}_{b)}f
+U^c\theta_{(a}\partial_{b)}(f+g)
+\half \left(\gamma^{ck}\partial_kf\right)\gammaU_{ab}\\[4pt]
+(f+g)\gamma^{ck}\theta_{(a}F_{b)k}
\label{deltaGamma}
\end{equation}
for all $a,b,c=0,\ldots,d$.

\goodbreak

Equation (\ref{deltaGamma}) is of central importance in our study; it yields the general form of the variations of the NC-connection compatible with the constraints $\nabla\gamma=0$, and $\nabla\theta=0$, and induced by the conformal Galilei re\-scalings~(\ref{deltagammadeltatheta}) and the Milne boosts~(\ref{deltaU}). 

\subsection{Conformal NC transformations: lightlike geodesics}
\label{NCCLightSection}

So far, we have been dealing with the Galilei-conformal symmetries of the equations of generic, i.e., timelike geodesics. What about those of the equations of \textit{lightlike geodesics} (\ref{spacelikegeod}) that model the worldlines of \textit{massless non-relativistic particles} \cite{Sou}?

Let us now determine the variations (\ref{deltaGamma}) of the NC-connection, $\Gamma$, that preserve the equations of lightlike geodesics, namely Equation (\ref{EqGeod}) supplemented by~$\dt=0$. 

We thus must have $\delta\Gamma^c_{ab}\dx^a\dx^b=\delta\mu\,\dx^c$, so that, necessarily, 
\begin{equation}
\gamma(df)=0
\label{gammadf=0}
\end{equation}
since $\theta_a\dx^a=0$ (the third term in the right-hand side of (\ref{deltaGamma}) has to vanish); this implies $df=f'\theta$, hence that $f$ is, along with $g$, a function of the time axis, $T$. We also find that $\delta\mu=0$. At last, the resulting variation of the NC-connection appears in the new guise
\begin{equation}
\delta\Gamma^c_{ab}
=
-f'\delta^c_{(a}\theta^{}_{b)}
+(f'+g')U^c\theta_a\theta_b
+
(f+g)\gamma^{ck}\theta_{(a}F_{b)k}
\label{deltaGammaLight}
\end{equation}
for all $a,b,c=0,\ldots,d$, where the unit vector field $U$, i.e.,  $\theta_aU^a=1$, and the two-form $F$ are as in (\ref{Gamma}).

We note that the constraint 
(\ref{fp+gp=0}), obtained in the ``massive'' case, does \textit{not} show up in the ``massless'' case. 

Just as in Section \ref{schSection}, we will assume that the variations (\ref{deltagammadeltatheta}) of the Galilei structure, and those (\ref{deltaGammaLight}) of the NC-connection, are generated by infini\-tesimal spacetime transformations.

\subsubsection{The conformal Newton-Cartan Lie algebra, $\cnc$, of null-pro\-jective conformal Galilei transformations}\label{cncSection}

The next natural step consists in demanding that the variations (\ref{deltagammadeltatheta}) of the Galilei structure, and those (\ref{deltaGammaLight}) of the NC-connection are, indeed, generated by infinitesimal spacetime transformations.

\goodbreak

We will thus define a new Lie algebra, called
the \textit{conformal Newton-Cartan Lie algebra},
and denoted by
$\cnc(M,\gamma,\theta,\Gamma)$, as the Lie algebra of those vector fields that are infinitesimal (i) conformal-Galilei transformation of $(M,\gamma,\theta)$, and (ii) transformations which permute \textit{lightlike} geodesics of $(M,\gamma,\theta,\Gamma)$. 
The \textit{conformal Newton-Cartan Lie algebra},  $\cnc(M,\gamma,\theta,\Gamma)$, is thus spanned by the vector fields, $X$, of $M$ such that
\begin{eqnarray}
\label{LXgamma}
L_X\gamma^{ab}&=&f\gamma^{ab}\\
\label{LXtheta}
L_X\theta_a&=&g\,\theta_a\\
\label{LXGamma}
L_X\Gamma^c_{ab}
\label{cnc} 
&=&
-f'\delta^c_{(a}\theta^{}_{b)}
+(f'+g')U^c\theta_a\theta_b
+
(f+g)\gamma^{ck}\theta_{(a}F_{b)k}
\end{eqnarray}
for all $a,b,c=0,1,\ldots,d$, where $f$ and $g$ are functions of the time axis, $T$, while $U$, and $F$ are as in (\ref{Gamma}) and (\ref{GammaU}).

\goodbreak

It is worth noticing that the Lie-transport (\ref{cnc}) of the NC-connection satisfies the very simple condition, viz., 
\begin{equation}
L_X\Gamma^{abc}=0
\label{Coriolis} 
\end{equation}
where $L_X\Gamma^{abc}=(L_X\Gamma^c_{k\ell})\gamma^{ak}\gamma^{b\ell}$. Interestingly, Equation (\ref{Coriolis}) is specific to the so-called \textit{Coriolis} Lie algebra of Galilei isometries of $(M,\gamma,\theta)$; see \cite{Duv2}.

Let us emphasize, at this stage, that the Schr\"odinger Lie algebra we have already  been dealing with in Section \ref{schzSection}, is clearly a Lie subalgebra of the conformal Newton-Cartan Lie algebra, viz.,
\begin{equation}
\sch_2(M,\gamma,\theta,\Gamma)
\subset
\cnc(M,\gamma,\theta,\Gamma)
\label{Coriolis2} 
\end{equation}
corresponding to the constraint
\begin{equation}
f+g=0
\label{f+g=0} 
\end{equation}
associated with the dynamical exponent $z=2$; see (\ref{z=2}). We will thus ignore, in the sequel, this special solution, and concentrate on the maximal solutions of Equations (\ref{LXgamma})--(\ref{LXGamma}) with $f+g\neq0$.

We will now determine the conformal Newton-Cartan Lie algebra in the flat case, i.e., the Lie algebra $\cnc(d)=\cnc(\bbR\times\bbR^d,\gamma,\theta,\Gamma)$ where $\gamma^{ab}$ and $\theta_a$ are as in~(\ref{flatNC}), as well as $\Gamma^c_{ab}=0$.

Let us put, in full generality, $U=\partial_0+U^A\partial_A$, where the $U^A$ are smooth functions of spacetime. 

The system (\ref{LXgamma})--(\ref{LXGamma}) to solve for $X=X^0\partial_0+X^A\partial_A$ reads then
\begin{eqnarray}
\label{1bis}
\partial_AX_B+\partial_BX_A&=&-f\delta_{AB}\\
\label{2bis}
\partial_AX^0&=&0\\
\label{3bis}
\partial_0X^0&=&g\\
\label{6bis}
\partial_0\partial_0X^A&=&(f'+g')U^A+(f+g)F_{0A}\\
\label{7bis}
\partial_0\partial_BX^A&=&-\half {}f'\delta^A_B-\half(f+g)F_{AB}\\
\label{9bis}
\partial_A\partial_BX^C&=&0
\end{eqnarray}
for all $A,B,C=1,\ldots,d$. 

Straightforward computation provides the general solution of that system. We find that $X^0=\xi(t)$, hence $g(t)=\xi'(t)$, remains arbitrary; Equations (\ref{1bis}), and~(\ref{9bis}) yield 
$X^A=\omega^A_B(t)x^B-\half {}f(t)x^A+\eta^A(t)$, the functions $\omega_{AB}(t)=-\omega_{BA}(t)$, $f(t)$, and $\eta^A(t)$ being unspecified.  Conspicuously, Equations~(\ref{6bis}), and (\ref{7bis}), bring no further restriction to the spatial components, $X^A$, as long as the two-form $F$, in the right-hand side of these equations, is not constrained whatsoever.
Indeed, we can easily deduce from (\ref{dotU}), and~(\ref{curlU}) that 
\begin{equation}
F_{AB}=2\partial_{[A}U_{B]}.
\label{FAB}
\end{equation}
and
\begin{equation}
F_{0A}=\partial_0U_A+U^B\partial_AU_B.
\label{F0A}
\end{equation}
Using then (\ref{7bis}), and (\ref{6bis}), we find that $\omega'_{AB}(t)x^B=(f+g)U_A+\partial_A\psi$, as well as $-\half{}f''(t)\delta_{AB}\,x^B+\eta''_A(t)=\partial_A\chi$, for some functions $\psi$, and $\chi$, and some unit vector field, $U$, of spacetime. Our claim is, hence, justified.

We contend that $X\in\cnc(d)$ iff
\begin{equation}
X=\xi(t)\frac{\partial}{\partial t}
+
\Big(\omega^A_B(t)x^B-\half {}f(t)x^A+\eta^A(t)\Big)\frac{\partial}{\partial x^A},
\label{cncd} 
\end{equation}
where, $\bomega(t)\in\so(d)$, $\veta(t)$, $\xi(t)$ and $f(t)$ depend smoothly on time, $t$, in an arbitrary fashion.

If $\cnc_z(d)$ denotes the Lie subalgebra with dynamical exponent $z$, i.e., defined by Equation (\ref{f+2g/z=0}), we trivially have
\begin{equation}
\cnc_z(d)\cong\cgal_z(d)
\label{cnczcongcgalz}
\end{equation}
in view of (\ref{cgalz}). Let us emphasize that dealing with \textit{rational} dynamical exponents, $z$, introduced in~(\ref{z}), is clearly allowed by the novel geometric definition (\ref{galconfMN}) of conformal Galilei transformations.

\subsubsection{The Lie algebra, $\cmil$, of null-pro\-jective conformal transformations of NC-Milne spacetime}

We will now confine considerations to the case where the NC-spacetime admits a prefer\-red geodesic and irrotational observer (an~``ether''), i.e., a unit vector field, $U$, such that, 
Condition (\ref{F=0}) holds true.
With the convention of Section \ref{NCGaugeSection}, we denote such a ``NC-Milne-structure'' by $(M,\gamma,\theta,\GammaU)$. 

Specializing the system (\ref{LXgamma})--(\ref{LXGamma}) to the case $F=0$, we thereby define the \textit{conformal Milne Lie algebra},  $\cmil(M,\gamma,\theta,\GammaU)$, as the \textit{maximal} Lie algebra of vector fields, $X$, of $M$ such that
\begin{eqnarray}
\label{LXgammaBis}
L_X\gamma^{ab}&=&f\gamma^{ab}\\
\label{LXthetaBis}
L_X\theta_a&=&g\,\theta_a\\
\label{LXGammaBis}
L_X\GammaU^c_{ab}
&=&
-f'\delta^c_{(a}\theta^{}_{b)}
+(f'+g')U^c\theta_{a}\theta_{b}
\end{eqnarray}
for all $a,b,c=0,1,\ldots,d$, where $f$ and $g$ are functions of the time axis, $T$.

We now determine the Lie algebra $\cmil(d)=\cmil(\bbR\times\bbR^d,\gamma,\theta,\GammaU)$, in the special case of flat NC-Milne spacetime specified by Equations (\ref{flatNC}), where $\Gamma^c_{ab}=\GammaU^c_{ab}=0$ for all $a,b,c=0,1,\ldots,d$, in a chosen inertial coordinate system --- we have, in particular, $U=\partial_0+U^A\partial_A$ where 
\begin{equation}
U^A=\const
\label{U=const}
\end{equation}
for all $A=1,\ldots,d$.
Indeed, an ether in flat NC-spacetime is a solution, $U$, of the~PDE (\ref{FAB}) and (\ref{F0A}) with $F=0$. We get $U_A=\partial_A\psi$ and $\partial_A(\partial_0\psi+\half{}U_BU^B)=0$. One can thus choose $\psi$ to be a solution of the free Hamilton-Jacobi equation
\begin{equation}
\partial_t\psi+\half\delta^{AB}\partial_A\psi\partial_B\psi=0
\label{HJ} 
\end{equation}
whose general solution, $\psi$, is well-known and leads to $\partial_A\psi=U_A$ where (\ref{U=const}) holds.

The system (\ref{LXgammaBis})--(\ref{LXGammaBis}) to solve for $X=X^0\partial_0+X^A\partial_A$ is now given by 
\begin{eqnarray}
\label{1ter}
\partial_AX_B+\partial_BX_A&=&-f\delta_{AB}\\
\label{2ter}
\partial_AX^0&=&0\\
\label{3ter}
\partial_0X^0&=&g\\
\label{6ter}
\partial_0\partial_0X^A&=&(f'+g')U^A\\
\label{7ter}
\partial_0\partial_BX^A&=&-\half {}f'\delta^A_B\\
\label{9ter}
\partial_A\partial_BX^C&=&0
\end{eqnarray}
for all $A,B,C=1,\ldots,d$. 

\goodbreak

Returning to the calculation done in Section \ref{cncSection}, we get $X^0=\xi(t)$, with $g(t)=\xi'(t)$, and 
$X^A=\omega^A_B(t)x^B-\half {}f(t)x^A+\eta^A(t)$, with the same notation as before. Equation~(\ref{7ter}) readily implies
\begin{equation}
\omega'_{AB}(t)=0
\label{omegapAB=0}
\end{equation}
while Equation (\ref{6ter}) yields
\begin{equation}
(f'+g')U^A=-\half{}f''\,x^A+(\eta^A)''.
\label{UA}
\end{equation}
This entails that $f''=0$, i.e.
\begin{equation}
f(t)=-2(\kappa{}t+\lambda).
\label{f(t)}
\end{equation}

\goodbreak

The latter equations therefore imply $(f'+g')U^A=(\eta^A)''$, leading us to the expression
\begin{equation}
\eta^A(t)=\alpha^A(\xi(t)-\kappa{}t^2)+\tbeta^A{}t+\tgamma^A
\label{etaA}
\end{equation}
where we have put
\begin{equation}
\alpha^A=U^A
\label{alpha=U}
\end{equation}
and where the coefficients $\tbeta^A$, and $\tgamma^A$ are integration constants. 

\goodbreak

Now,  if $X_1,X_2$ are solutions of the system (\ref{LXgammaBis})--(\ref{LXGammaBis}), so is their Lie bracket $X_{12}=[X_1,X_2]$. This yields the consistency relation $f_{12}=X_1f_2-X_2f_1$ which reads here $\kappa_{12}t+\lambda_{12}=\xi_1(t)\kappa_2-\xi_2(t)\kappa_1$. Thus $\xi(t)=\kappa\,u(t)+\mu{}t+\varepsilon$, for some function~$u$, the coefficients $\mu$, and $\varepsilon$ being constant.
Exploiting the fact that $X\mapsto{}X^0\partial_0$ is a Lie algebra homomorphism into $\Vect(\bbR)$, we write $\xi_{12}(t)=\xi_1(t)\xi'_2(t)-\xi_2(t)\xi'_1(t)$; straightforward calculation then shows that $u$ is a polynomial of degree $2$, which, up to lower degree terms, is given by
\begin{equation}
u(t)=\half{}c\,t^2
\label{u}
\end{equation}
with 
\begin{equation}
(c-1)(c-2)=0.
\label{c-constraint}
\end{equation}
We thus find $X^0=\xi(t)$ where $\xi(t)=\half{}\kappa{}c\,t^2+\mu{}t+\varepsilon$, and $X^A=\omega^A_B\,x^B+\kappa{}t{}x^A+\lambda{}x^A+\half(c-2)\kappa t^2\alpha^A+\beta^A{}t+\gamma^A$, with $\beta^A$, and $\gamma^A$ new integration constants.

The case $c=2$ gives back the expanded Schr\"odinger Lie algebra, $\widetilde{\sch}(d)$, see~(\ref{schd}), already studied since $f'(t)+g'(t)=0$. 
 It is not the only possibility, though.\footnote{The Schr\"odinger Lie algebra is the Lie algebra of a group of spacetime transformations that actually permute all geodesics, in particular lightlike geodesics.}

Consider then the new case $c=1$. We claim that $X\in\cmil(d)$ iff
\begin{eqnarray}
X
&=&
\nonumber
\left(\half{}\kappa{}t^2+\mu{}t+\varepsilon\right)\frac{\partial}{\partial t}\\[6pt]
&&+
\left(\omega^A_B\,x^B+\kappa{}t{}x^A+\lambda{}x^A-\half{}t^2\alpha^A+\beta^A t+\gamma^A\right)\frac{\partial}{\partial x^A}
\label{cmild} 
\end{eqnarray}
where $\bomega\in\so(d)$, $\balpha,\bbeta,\bgamma,\in\bbR^d$, and $\kappa,\lambda,\mu,\varepsilon\in\bbR$. 

Let us highlight that, just as in the case of $\widetilde{\sch}(d)$, time and space dilations are independent within $\cmil(d)$. As for the parameter, $\balpha$, in (\ref{cmild}) it serves as a novel \textit{acceleration} generator \cite{LSZ, LSZGalconf}.

\subsubsection{
Conformal NC-Milne Lie algebras, $\cmil_z$, with dynamical exponent~$z$; the CGA 
Lie algebra}\label{CGAzSection}

Much in the same way than in Section \ref{cgalzSection}, we will now introduce subalgebras of the conformal NC-Milne Lie algebra with prescribed dynamical exponent, $z$.

We will define $\cmil_z(M,\gamma,\theta,\GammaU)$ as the Lie subalgebra of $\cmil(M,\gamma,\theta,\GammaU)$ defined by Equation (\ref{f+2g/z=0}); we will call it the \textit{conformal NC-Milne Lie algebra with dynamical exponent $z$}.

Let us lastly establish, in the case of a flat NC-Milne structure, the expression of the generators of the Lie algebra $\cmil_z(d)=\cmil_z(\bbR\times\bbR^d,\gamma,\theta,\GammaU)$. Those retain the form (\ref{cmild}) where, in view of (\ref{3ter}), and (\ref{f(t)}), Equation (\ref{f+2g/z=0}) writes
\begin{equation}
\frac{1}{z}\left((1-z)\kappa{}t+(\mu-z\lambda)\right)=0.
\label{f+2g/z=0Bis}
\end{equation}

$\bullet$ In the generic case, $f'\neq0$, one ends up with
\begin{equation}
z=1
\label{z=1}
\end{equation}
and
\begin{equation}
\mu=\lambda
\label{mu=lambda}
\end{equation}
which entails that time and space are related in the same way.
Therefore $\cmil_1(d)$ is spanned by the vector fields (\ref{cmild}) for which (\ref{mu=lambda}) holds. It is isomorphic to the CGA, namely, the \textit{Conformal Galilean Algebra} (\ref{CGA}) of Lukierski, Stichel and Zakrzewski  \cite{LSZGalconf}, i.e.,
\begin{equation}
X=\Big(\half \kappa t^2+\lambda t+\varepsilon\Big)\frac{\partial}{\partial t}
+
\Big(
\omega^A_B\,x^B+\lambda{}x^A+\kappa{}t{}x^A-\half \alpha^A{}t^2+\beta^At+\gamma^A\Big)\frac{\partial}{\partial x^A}.
\label{CGAbis} 
\end{equation}
where $\bomega\in\so(d)$, $\balpha,\bbeta,\bgamma,\in\bbR^d$, and $\kappa,\lambda,\varepsilon\in\bbR$. 

The Lie algebra $\cmil_1(d)$ admits the faithful $(d+3)$-dimensional representation $X\mapsto{}Z$, where
\begin{equation}
Z=
\left(
\begin{array}{cccc}
\bomega&-\half\balpha&\bbeta&\bgamma\\[6pt]
0&\lambda&2\varepsilon&0\\[6pt]
0&\half\kappa&0&\varepsilon\\[6pt]
0&0&-\kappa&-\lambda
\end{array}
\right)
\label{CGARep} 
\end{equation}
with the same notation as before. 

Again, the following decomposition holds
\begin{equation}
\cmil_1(d)\cong(\so(d)\times\so(2,1))\ltimes(\bbR^d\times\bbR^d\times\bbR^d).
\label{CGALevyMalcev} 
\end{equation}

$\bullet$ In the special case, $f'=0$, i.e., $\kappa=0$
(no expansions), and $\mu=z\lambda$, we discover a whole family of Lie subalgebras $\cmil_z(d)\subset\cmil(d)$ parametrized by an arbitrary dynamical exponent, $z$. 

In the limit $z\to\infty$, where $f=0$, the Lie subalgebra $\cmil_\infty(d)\subset\cmil(d)$ is spanned by the vector fields (\ref{cmild}) with $\kappa=\lambda=0$. Let us stress that, in this limiting case, space dilations are ruled out (compare Equation (\ref{zinfinity})).

In both cases we obtain the Lie algebra of vector fields
\begin{equation}
X=\Big(\mu t+\varepsilon\Big)\frac{\partial}{\partial t}
+
\Big(
\omega^A_B\,x^B+\frac{\mu}{z}\,x^A-\half \alpha^A{}t^2+\beta^At+\gamma^A\Big)\frac{\partial}{\partial x^A}.
\label{genzCGA} 
\end{equation}
with the same notation as before.

\subsubsection{
The finite-dimensional conformal Galilei Lie algebras, $\alt_{2/N}(d)$
}\label{altSection}

Our formalism leads thus to an intrinsic definition of distinguished finite-dimensional subalgebras of the conformal Galilei Lie algebra $\cgal(d)$, namely $\sch_2(d)$, and $\cmil_1(d)$ with dynamical exponents $z=2$, and $z=1$ respectively (see~(\ref{sch2dLeviMalcev}), and (\ref{CGALevyMalcev})). Restricting, here, considerations to the very special case of \textit{flat} NC structures (expres\-sed in a given Cartesian coordinate system), one might search for other finite-dimensional Lie subalgebras of the conformal Galilei Lie algebras, $\cgal_z(d)\cong\cnc_z(d)$, with prescribed dynamical exponent $z$; see (\ref{cnczcongcgalz}). 

Recall (see (\ref{cgalz})) that $\cgal_z(d)$ is generated by those $X\in\Vect(\bbR\times\bbR^d)$ of the form
\begin{equation}
X=\xi(t)\frac{\partial}{\partial t}
+
\Big(\omega^A_B(t)x^B+\frac{1}{z}\xi'(t)x^A+\eta^A(t)\Big)\frac{\partial}{\partial x^A}
\label{cgalzBis} 
\end{equation}
where $\bomega(t)\in\so(d)$, $\veta(t)$, and $\xi(t)$ depend \textit{smoothly} on on time, $t$. 

Previous experience with the above-mentioned Lie algebras prompts us to look for Lie algebras of \textit{polynomial} --- not merely smooth --- vector fields of $\cgal_z(d)$. 

Consider, hence, vector fields, $X\in\cgal_z(d)$, that are polynomials of fixed degree $N>0$ in the variables $t=x^0,x^1,\ldots,x^d$. This entails the following decompositions $\bomega(t)=\sum_{n=0}^N{\omega_n t^n}$, $\veta(t)=\sum_{n=0}^N{\veta_n t^n}$, and $\xi(t)=\sum_{n=0}^N{\xi_n t^n}$, since the spatial components $X^A$ are already of first order in $x^1,\ldots,x^d$. Bearing in mind that $X\mapsto\xi$ is a Lie algebra homomorphism, we claim that the $\xi=\xi(t)\partial/\partial t$ do span a polynomial Lie subalgebra of $\Vect(\bbR)$, hence a Lie subalgebra of $\Sl(2,\bbR)$ since the latter is maximal in the Lie algebra, $\Vect^\mathrm{Pol}(\bbR)$, of polynomial vector fields of~$\bbR$. We therefore find
$\xi_n=0$ for all $n\geq3$, so that 
\begin{equation}
\xi(t)=\half\kappa t^2+\mu t+\varepsilon
\label{sl2} 
\end{equation}
with $\kappa,\mu,\varepsilon\in\bbR$.

Let now us seek under which condition (if any) the Lie bracket $X_{12}=[X_1,X_2]$ of two such polynomial vector fields $X_1$ and $X_2$ is, itself, polynomial of degree $N$, Condition (\ref{sl2}) being granted. Straightforward calculation yields
\begin{eqnarray}
\label{Eqxi}
\xi_{12}&=&\xi_1\xi'_2-\xi_2\xi'_1\\
\label{Eqomega}
\bomega_{12}&=&[\bomega_2,\bomega_1]+\xi_1\bomega'_2-\xi_1\bomega'_1\\
\label{Eqoeta}
\veta_{12}&=&\bomega_2\veta_1-\bomega_1\veta_2+\xi_1\veta'_2-\xi_2\veta'_1-\frac{1}{z}\left(\xi'_1\veta_2-\xi'_2\veta_1\right).
\end{eqnarray}
Condition (\ref{Eqxi}) brings no further restriction in view of (\ref{sl2}). From (\ref{Eqomega}), we discover that, necessarily, $\bomega'_1=\bomega'_2=0$; this entails that 
\begin{equation}
\bomega\in\so(d)
\label{sod} 
\end{equation}
in (\ref{cgalzBis}). At last, we readily find that the right hand-side of Equation (\ref{Eqoeta}) turns out to be a polynomial of degree $N+1$ in $t$, namely $\veta_{12}=\sum_{n=0}^{N+1}{(\veta_{12})_nt^n}$ with
$(\veta_{12})_{N+1}=\left(\half{}N-z\right)(\kappa_1(\veta_2)_N-\kappa_2(\veta_1)_N)$. In order to acquire a Lie algebra of polynomial vector fields of degree $N>0$, we must simply impose the constraint
\begin{equation}
z=\frac{2}{N}
\label{z=2/N}
\end{equation}
on the dynamical exponent. At last, we have shown that, in Equation (\ref{cgalzBis}),
\begin{equation}
\veta(t)=\veta_N t^N+\cdots+\veta_1 t+\veta_0
\label{vetat}
\end{equation}
with $\veta_n\in\bbR^d$ for all $n=0,1,\ldots,N=2/z$.

\goodbreak

We claim that the finite-dimensional Lie subalgebras of $\cgal_{2/N}(d)$ defined by (\ref{sl2}), (\ref{sod}), and (\ref{vetat}) together with (\ref{z=2/N}) are isomorphic with the so-called $\alt_{2/N}(d)$ Lie algebras discovered by Henkel \cite{Hen} in the study of scale invariance for strongly anisotropic critical systems (with $d=1$).\footnote{The definition of these Lie algebras clearly involves constraints given by differential operators of higher order, which go, strictly speaking, beyond our formalism relying essentially on \textit{second order} PDE associated with transport equations of NC-structures.} We have thus proved that
\begin{equation}
\cgal_{2/N}^\mathrm{Pol}(d)\cong\alt_{2/N}(d).
\label{altd}
\end{equation}

Note the special cases $\cgal_{2}^\mathrm{Pol}(d)=\sch_2(d)$, and $\cgal_{1}^\mathrm{Pol}(d)=\cmil_1(d)$ corresponding to $N=1$, and $N=2$ respectively.

It would be desirable to find a truly geometric definition of such Lie subalgebras of the Lie algebra of conformal Galilei Lie algebras, $\cgal(M,\gamma,\theta)$, in the case of an arbitrary Galilei (or Newton-Cartan) structure.

\section{Conformal Galilean symmetries of physical systems}\label{dynSys}

In order to illustrate our general formalism, we first present a framework, due original\-ly to Souriau \cite{Sou}, which allows us to describe, in particular, both massive and massless Galilean elementary systems in a unified way.

\goodbreak

Consider a Hamiltonian system with $d$ degrees of freedom, whose \textit{phase space} is a $2d$-dimensional symplectic manifold $(\cM,\Omega)$, and whose Hamiltonian is a smooth function, $H$, of $\cM$. The two-form $\Omega=\half\Omega_{\alpha\beta}\,d\xx^\alpha\wedge{}d\xx^\beta$ of~$\cM$ is closed, $d\Omega=0$, and non-degenerate, $\det(\Omega_{\alpha\beta})\neq0$. Using its inverse, $\Omega^{-1}=\half\Omega^{\alpha\beta}\partial_{\alpha}\wedge\partial_{\beta}$, we get the Poisson bracket $\{F,G\}=\Omega^{\alpha\beta}\p_\alpha{F}\,\p_\beta{G}$ of two observables $F$, and $G$. (The Jacobi identity is equivalent to $d\Omega=0$.) Then Hamilton's equations read
\begin{equation}
\frac{d\xx^\alpha}{dt\ }=\{H,\xx^\alpha\}
\label{Hameq}
\end{equation}
where $\alpha=1,\dots,2d$, the parameter $t$ being interpreted as ``time''. 

If $X_H=\{H,\,\cdot\,\}$ is the associated Hamiltonian vector field, we see that (\ref{Hameq}) can also be written as\footnote{
The form (\ref{varform}) of Hamilton's equations allows for a variational interpretation; see, e.g., \cite{Sou,HPALandau,Tuy}.} 
\begin{equation}
\frac{d\xx}{dt}=X_H
\qquad
\hbox{where}
\qquad
\Omega(X_H)=-dH
\label{varform}
\end{equation}
or, using coordinates, $\Omega_{\alpha\beta}\,X^\alpha_H=-\partial_\beta{H}$, for all $\beta=1,\ldots,2d$.

\goodbreak

One can go one step farther \cite{AM} and, promoting time as a new coordinate, consider the $(2d+1)$-dimensional ``evolution space'' $\cV=\cM\times\bbR$ endowed with the following closed two-form which we write, with some abuse of notation, as
\begin{equation}
\sigma=\Omega-dH\wedge{}dt.
\label{Ssigma}
\end{equation}
The vector fields $Y=\lambda(X_H+\partial/\partial{}t)$ of $\cV$, with $\lambda\in\bbR$, clearly satisfy $\sigma(Y)=0$ since $X_H(H)=0$, and $Yt=\lambda$. Denoting $y=(\xx,t)$ points of $\cV$, we see that the equations of motion~(\ref{Hameq}) admit the alternative form
\begin{equation}
\frac{dy}{d\tau}=Y
\qquad
\hbox{with}
\qquad
\sigma(Y)=0
\label{kersigma}
\end{equation}
where $\tau$ is now an arbitrary curve-parameter.

\goodbreak

Conversely, let us consider a closed two-form, $\sigma$, on some general \textit{evolution space}, $\cV$, whose kernel, $K=\ker(\sigma)$, has a (nonzero) constant dimension.\footnote{One says that the two-form is \textit{presymplectic}.}
Then~$K$ (see (\ref{kersigma})) is an integrable distribution. So, there exists, passing through each point~$y\in\cV$, a submanifold whose tangent space is spanned by 
those vectors in~$K$. Each leaf (or characteristic) of $K$ is a \textit{classical motion}. The set of these motions (which is assumed to be a well-behaved manifold) is Souriau's \textit{space of motions}, $\cU=\cV/K$, of the system. 
The two-form $\sigma$ passes to the quotient, $\cU$, which becomes, hence, a symplectic manifold. Espousing this point of view, one regards the evolution space as fundamental since it hosts the dynamics in a purely intrinsic way. See also the recent essay \cite{Rov} supporting this standpoint in the classical and quantum context. 

A \textit{symmetry} of the evolution space $(\cV,\sigma)$ is given by a vector field $Z$ which Lie-transports the two-form $\sigma$, namely such that
\begin{equation}
L_Z\sigma=0.
\label{Ssymm}
\end{equation}
A symmetry is called \textit{Hamiltonian} if there exists a function $\cJ_Z$ of $\cV$ such that, globally,
\begin{equation}
\sigma(Z)=-d\cJ_Z.
\label{presympNoether}
\end{equation}
Then, one readily finds that $Y\cJ_Z=0$ for all $Y\in{}K$, i.e., that $\cJ_Z$ (determined by~(\ref{presympNoether}) up to an overall constant) is a conserved quantity. See \cite{Sou} for an account on this formulation of Noether's theorem.

Conversely, symplectic manifolds upon which a given group of Hamiltonian symmetries acts transitively can be constructed in a systematic fashion \cite{Sou}. For example, the homogeneous symplectic manifolds of the Galilei group will represent the spaces of motions of classical, non-relativistic, elementary particles. Skipping the details, here we simply list the results which are important for our purposes.

\subsection{Galilean massive particles} 
\label{Gal3msSection}

Generic elementary systems of the Galilei group (whose Lie algebra has been defined in (\ref{gal})) in four-dimensional, flat, NC-spacetime are classified by the \textit{mass}, $m$, and \textit{spin}, $s$, invariants. In the ``massive'' case, $m>0$, the evolution space of a spinning particle, with $s>0$, is $\cV=\bbR\times\bbR^3\times\bbR^3\times{}S^2$ parametrized by the quadruples $y=(t,\bx,\bv,\bu)$, and endowed with the two-form
\begin{equation}
\sigma=m\,dv_A\wedge(dx^A-v^A dt)-\frac{s}{2}\epsilon_{ABC}\,u^Adu^B\wedge{}du^C
\label{sigmams}
\end{equation}
where $\epsilon_{ABC}$ is the Levi-Civita symbol with $\epsilon_{123}=1$. 

\goodbreak

Equation~(\ref{sigmams}) happens to be of the form (\ref{Ssigma}) that unifies the symplectic structure of phase space $\cM=\bbR^3\times\bbR^3\times{}S^2$ and the Hamiltonian, namely
\begin{eqnarray}
\sigma=dp_A\wedge{}dx^A-\frac{1}{2s^2}\epsilon_{ABC}\,s^Ads^B\wedge{}ds^C
-d\left(\frac{\bp^2}{2m}\right)\wedge{}dt
\label{msSympHam}
\end{eqnarray}
where the vector $\bp=m\bv$ stands for the linear momentum, and $\bs=s\bu$ for the \textit{clas\-sical spin}.\footnote{At the  purely classical level studied here, $s$ is an arbitrary positive number; the ``pre\-quantizability'' \cite{Sou} requires it to be a half-integral multiple of $\hbar$.}
Ordinary phase space has been extended by the sphere $S^2$, endowed with its canonical surface element. 
The two-form (\ref{sigmams}) is closed and has a one-dimensional kernel; the characteristic curves, which are solutions of the
free equations of motion~(\ref{kersigma}), namely
\begin{equation}
\dt=1,
\qquad
\dot\bx=\bv,
\qquad
\dot{\bv}=0,
\qquad
\dot{\bu}=0
\label{mseqmot}
\end{equation}
project on spacetime as usual straight worldlines. Those are independent of spin, which is itself a constant of the motion. 

\goodbreak

These worldlines are, in fact, timelike geodesics since
\begin{equation}
\dot{t}\neq0.
\label{timelike}
\end{equation}

As for the symmetries of the model coming from conformal Galilean transformations of flat spacetime, one shows \cite{Duv0,Duv1} that the only vector fields $X\in\cgal(3)$ that admit a lift, $\wX$, to $(\cV,\sigma)$ verifying $L_\wX\sigma=0$ (see (\ref{Ssymm})) are necessarily Schr\"odinger vector fields, $X\in\sch(3)$. 
The explicit expression is
\begin{eqnarray}
\wX
&=&
\big(\kappa{}t^2+2\lambda{}t+\varepsilon\big)\frac{\partial}{\partial t}
+
\big(\omega^A_B\,x^B+\kappa{}t{}x^A+\lambda{}x^A+\beta^A{}t+\gamma^A\big)\frac{\partial}{\partial{}x^A}\nonumber
\\
&&
\label{mspimp}
+\big(\omega^A_B\,v^B+\beta^A-\lambda{}v^A
+\kappa(x^A-v^At)\big)
\frac{\partial}{\partial{}v^A}\\
\nonumber
&&+\omega^A_B\,u^B\frac{\partial}{\partial{}u^A}
\end{eqnarray}
with the notation of (\ref{sch2d}). 

\goodbreak

We, hence, recover the results of Section \ref{schSection} dealing with the symmetries of the equations of timelike NC-geodesics. Notice that the Schr\"odinger symmetry still holds in the presence of spin.

\goodbreak

The action (\ref{mspimp}) is Hamiltonian, and the conserved quantities, calculated using Noether's theorem (see Equation (\ref{presympNoether})) read\footnote{We have put $\omega_A=-\half\epsilon_{ABC}\omega^{BC}$ for all $A=1,2,3$.}
\begin{equation}
\cJ_\wX=\bJ\cdot\bomega-\bG\cdot\bbeta+\bP\cdot\bgamma-H\varepsilon-K\kappa+D\lambda
\label{Jm}
\end{equation}
where
\begin{equation}
\begin{array}{llll}
\bP&=&\bp&\hbox{\small Linear momentum}
\\[12pt]
\bg&=&m\bq&\hbox{\small Galilean boost}
\\[12pt]
\bJ&=&\bx\times\bp+s\bu&\hbox{\small Angular momentum}
\\[12pt]
H&=&\displaystyle\frac{\bp^2}{2m}&\hbox{\small Energy}
\\[12pt]
K&=&\displaystyle\frac{m\bq^2}{2}&\hbox{\small Schr\"odinger expansions}
\\[12pt]
D&=&\bp\cdot\bq&\hbox{\small Schr\"odinger dilations}
\end{array}
\label{msConsQuant}
\end{equation}
together with
\begin{equation}
\bq=\bx-\bv{}t.
\label{q}
\end{equation}

\goodbreak

Note that the spin enters the angular momentum only
and is, in fact, separately conserved. The space of motions, $\cU=\bbR^3\times\bbR^3\times{}S^2$, therefore inherits from $\sigma$ the symplectic two-form $\Omega=dp_A\wedge{}dq^A-(s/2)\epsilon_{ABC}\,u^Adu^B\wedge{}du^C$. The associated Poisson brackets of the components (\ref{msConsQuant}) of the \textit{moment map} \cite{Sou} then realize the
\textit{one-parameter central extension} of the  Schr\"odinger group, via
\begin{equation}
\{P_A,G_B\}=m\,\delta_{AB}.
\label{mcentext}
\end{equation}

As a further  example of massive Schrodinger symmetry, we mention
non-relativis\-tic Chern-Simons vortices \cite{CSvort}.

So far, we have only studied free particles.
Let us mention that the $\so(2,1)$
symmetry would survive, if $d=3$, the addition of
a Dirac monopole \cite{Jmon,Duv0,HPAmon}, and, if
$d=2$, that of a ``magnetic vortex''
\cite{Jvort,DHvort}.

\subsection{Galilean symmetry in hydrodynamics}\label{hydrosection}

Another example with Schr\"odinger symmetry involves
hydrodynamics \cite{fluid,JackiwFluid,Bazeia}. 
To shed a new light on the problem, we present our
results in a way complementary to the geometric approach
followed in the previous sections.

The equations of motion of an isentropic and dissipationless fluid, in flat $(d+1)$-dimensional non-relativistic spacetime, read \cite{JackiwFluid}
\begin{eqnarray}
\partial_t\rho+\bnabla\cdot(\rho\,\bv)&=&0,
\label{continuity}
\\[8pt]
\partial_t\bv+\bv\cdot\bnabla\bv&=&-\,\bnabla V'(\rho),
\label{Eulereq}
\end{eqnarray}
where $\rho(t,\bx)$ is the density and $\bv(t,\bx)$ the velocity field.

The enthalpy, $V'(\rho)$, is related to the pressure, $P$, via $\rho V'(\rho)-V(\rho)=P$. For simplicity, we focus our attention to the irrotational case, 
$\bv=\bnabla\theta$, ---
where~Ê$\theta(t,\bx)$ is a potential for the velocity field --- when the system can be derived from a variational principle using the Lagrangian
\begin{equation}
L=L_0-V(\rho)=-\rho\big(\p_t\theta+\half(\bnabla\theta)^2\big)-V(\rho).
\label{irrotLag}
\end{equation}
Varying $L$ in (\ref{irrotLag}) with respect to $\theta$ yields the continuity equation
(\ref{continuity}), and varying it with respect to $\rho$ yields the
Bernoulli equation
\begin{eqnarray}
\p_t\theta+\half(\bnabla\theta)^2=-V'(\rho)
\label{Bernoulli}
\end{eqnarray}
whose gradient is the Euler equation (\ref{Eulereq}).

\goodbreak

The system is plainly Galilei-invariant: a boost
implemented by $\theta\mapsto\theta^*$, and $\rho\mapsto\rho^*$, where 
\begin{eqnarray}
\theta^*(t,\bx)&=&\theta(t,\bx+\bb t)-\bb\cdot\bx-\half\bb^2t
\label{boostimptheta}
\\
\rho^*(t,\bx)&=&\rho(t,\bx+\bb{}t)
\label{boostimprho}
\end{eqnarray}
leaves the Lagrangian (\ref{irrotLag}) invariant. 
Routine calculation proves the invariance against space and time translations, as well as rotations \cite{fluid,JackiwFluid}, proving the
full Galilean invariance of the model.
 
Now we inquire about the conformal symmetries.

\medskip\noindent
\textbf{Scale invariance}

$\bullet$ Consider a dilation with dynamical exponent, $z$, namely
\begin{equation}
t^*=\lambda^zt,
\qquad
\bx^*=\lambda\,\bx
\label{zdil}
\end{equation}
and attempt to implement it as
$$
\theta^*=\lambda^a\theta(t^*,\bx^*),
\qquad
\rho^*=\lambda^b\rho(t^*,\bx^*)
$$
where $a$ and $b$ have to be determined. 

\goodbreak

The two terms  in the free Lagrangian, $L_0$, are seen to scale in the same way when
$a=z-2,$
and then the entire expression scales by
$
\lambda^{b+z-2}.
$
However, the measure of integration scales as
$dtd\bx=\lambda^{-(d+z)}dt^*d\bx^*$, where $d\bx=dx^1\ldots{}dx^d$.
Invariance of the free Lagrangian density requires therefore
$
b=d+(2-z).
$
Thus, for \textit{any dynamical exponent}, $z$,
the free Lagrangian density is scale-invariant, whenever
\begin{eqnarray}
\theta^*(t,\bx)&=&\lambda^{z-2}\theta(t^*,\bx^*)
\label{zthetaimp}
\\
\rho^*(t,\bx)&=&\lambda^{d-z+2}\rho(t^*,\bx^*).
\label{zrhoimp}
\end{eqnarray}

Which potential can be added? Restricting ourselves to
the polytropic expres\-sion $V(\rho)=c\rho^\gamma$, we find
$
V(\rho^*)=\lambda^{\gamma(d-z+2)}V(\rho),
$
and also $V(\rho^*)=\lambda^{d+z}V(\rho)$ to match the free case. Therefore, to preserve the symmetry with respect to (\ref{zdil}),
the polytropic exponent must be
\begin{equation}
\gamma=\frac{d+z}{d+2-z}\,.
\label{zgamma}
\end{equation}  
Conversely, to deal with a potential $V(\rho)=c\rho^\gamma$ having dilations as symmetries requires to
choosing the dynamical exponent as
\begin{equation}
z=\frac{\gamma(d+2)-d}{\gamma+1}\,.
\label{zfixed}
\end{equation}
In other words, the potential breaks to (\ref{zfixed})
the freedom of choosing $z$.

\goodbreak

$\bullet$ For $z=2$, in particular,  when time is twice-dilated with respect to space,
\begin{equation}
t^*=\lambda^2\,t,
\qquad
\bx^*=\lambda\,\bx
\end{equation}
we recover the known results \cite{fluid}
\begin{eqnarray}
\theta^*(t,\bx)&=&\theta(t^*,\bx^*)
\label{Schrdilthetaimp}
\\
\rho^*(t,\bx)&=&\lambda^{d}\rho(t^*,\bx^*)
\label{Schrdilrhoimp}
\\[6pt]
\gamma&=&1+\displaystyle\frac{2}{d}\,.
\label{Schrdilexponent}
\end{eqnarray}

\goodbreak

\medskip\noindent
\textbf{Expansions}

Schr\"odinger expansions, viz.,
\begin{equation}
t^*=\displaystyle\frac{t}{1-\kappa t},\qquad
\bx^*=\displaystyle\frac{\bx}{1-\kappa t}
\label{z2expan}
\end{equation}
implemented as
\begin{eqnarray}
\rho^*(t,\bx)&=&(1-\kappa t)^{-d}
\rho(t^*,\bx^*)
\label{rhoexpimp}
\\
\theta^*(t,\bx)&=&\theta(t^*,\bx^*)-
\frac{\kappa\,\bx^2}{2(1-\kappa t)}
\label{thetaexpimp}
\end{eqnarray}
are readily seen to be symmetries for the free 
fluid system. 

\goodbreak

Let us attempt to generalize (\ref{z2expan})  as
\begin{equation}
t^*=\Omega\,t,\qquad
\bx^*=\Omega^\alpha\,\bx
\label{genexpan}
\end{equation}
and (\ref{rhoexpimp}), (\ref{thetaexpimp}) as
\begin{eqnarray}
\rho^*(t,\bx)&=&\Omega^{\delta}\rho(t^*,\bx^*)
\label{rhoexpimp2}
\\
\theta^*(t,\bx)&=&\theta(t^*,\bx^*)-
\beta\, \kappa\,\Omega^\gamma\, {\bx^*}^2
\label{thetaexpimp2}
\end{eqnarray}
where $\Omega=(1-\kappa t)^{-1}$,
and $\alpha,\beta,\gamma,\,\delta$ are to be determined.
Then the $\theta$-part of the free Lagrangian transforms according to
\begin{eqnarray}
\nonumber
\partial_t\theta^*+\half(\bnabla\theta^*)^2&=&
\Omega^2\partial_{t^*}\theta+\half\Omega^{2\alpha}(\bnabla^*\theta)^2\\[6pt]
&&
+\kappa\,\bx^*\cdot\bnabla^*\theta\left(\alpha\Omega-2\beta\Omega^{2\alpha+\gamma}\right)\\[6pt]
&&+
\nonumber
\beta\kappa^2(\bx^*)^2\left(2\beta\Omega^{2\alpha+2\gamma}-(2\alpha+\gamma)\Omega^{\gamma+1}\right).
\end{eqnarray}
Getting a symmetry requires, therefore,
$\alpha=1, \beta=\half, \gamma=-1$, and $\delta=d$,
leading to the above expressions (\ref{rhoexpimp}), and (\ref{thetaexpimp}).
This is, hence, the only case allowed by the expansion-symmetry in fluid mechanics.

On the other hand, dilations and Schr\"odinger
expansions generate, along with time translations, the (neutral component of the) group $\SO(2,1)$, only when the dynamical exponent is $z=2$.
The only consistent way to combine dilations and expansions is, hence, 
when the system carries a full Schr\"odinger symmetry \cite{fluid,JackiwFluid}. 

\goodbreak

\medskip\noindent
\textbf{
Conserved quantities}

Noether's theorem associates conserved quantities to
symmetries. In the present field-theoretic context, it goes as follows.
Let $\phi$ be any field.
An infinitesimal transformation, $\delta\phi$, 
is a symmetry if it changes the Lagrange density by a ``surface term'',
$\delta{ L}=\p_aC^a$,
for some quantities $C^a$. Then 
\begin{equation}
J^a=\frac{\delta L}{\delta(\p_a\phi)}\delta\phi-C^a
\end{equation} 
is a conserved current, $\p_{\alpha}J^\alpha=0$, so that the integral
\begin{equation}
Q=\displaystyle{\int_{t=t_0}{\!\!\!\!\!d\bx\,
{\Big(\frac{\delta L}{\delta(\p_{t}\phi)}\delta\phi-C^t\Big)}}}
\label{consquant}
\end{equation}
is a constant of the motion, i.e., is independent of $t_0$.

Returning to the Schr\"odinger case, and
using the Noether theorem one finds
the conserved quantities,
\begin{equation}
\begin{array}{llllc}
\bP&=&
\displaystyle\int\!d\bx\,\rho\,\bnabla\theta
&\hbox{\small Linear momentum}
\\[10pt]
\bG&=&\displaystyle\int\! d\bx\, \rho\big( 
\bx-\bnabla\theta\, t\big)
&\hbox{\small Galilean boosts}
\\[10pt]
\bJ&=&\displaystyle\int{\! d\bx\,\rho\, 
\bx\wedge\bnabla\theta}
&\hbox{\small Angular momentum}
\\[10pt]
H&=&\displaystyle\int{\!d\bx
\left(
\frac{1}{2}\rho(\bnabla\theta)^2+V(\rho)\right)}
&\hbox{\small Energy}
\\[10pt]
K&=&-t^2H+2tD
+\displaystyle\frac{1}{2}\!\displaystyle\int{\!d\bx\,
\rho\,\bx^2}\qquad
&\hbox{\small Schr\"odinger expansion}
\\[10pt]D&=&tH
-\displaystyle\frac{1}{2}\displaystyle\int\! d\bx\,\rho\,(\bx\cdot\bnabla\theta),
&\hbox{\small Schr\"odinger  dilation}
\\[10pt]
M&=&\displaystyle\int\! d\bx\,\rho\qquad\hfill
&\hbox{\small Mass}\\
\end{array}
\label{fluidSchr}
\end{equation}
where we have also added the total Galilean mass. Under (suitably defined) Poisson brackets, we get the generators of the \textit{one-parameter centrally extended Schr\"odinger algebra}~\cite{fluid}.
  
In conclusion, the free system admits, for any $z\neq2$, the expansion-less and dilations-only Lie subalgebra of $\sch_z(d)$ as a Lie algebra of symmetries. It 
 possesses the full Schr\"odinger Lie algebra of symmetries (including
expansions) when $z=2$.

The symmetry is preserved when the polytropic exponent
is chosen suitably, namely as in (\ref{Schrdilexponent}).

\goodbreak

\medskip\noindent
\textbf{Accelerations} 

It is worth mentioning that accelerations,
\begin{equation}
t^*=t,
\qquad
\bx^*=\bx-\half\ba t^2,
\label{acceleration}
\end{equation}
implemented as
\begin{eqnarray}
\theta^*(t,\bx)=\theta(t^*,\bx^*)+(\ba\cdot\bx^*)t^*,
\qquad
\rho^*(t,\bx)=\rho(t^*,\bx^*)
\label{thetaaccelimp}
\end{eqnarray}
change the Lagrangian as,
\begin{equation}
\rho^*\Big(\p_{t}\theta^*+\half(\bnabla\theta^*)^2
\Big)=
\rho\Big(\p_{t^*}\theta+\half(\bnabla^*\theta)^2\Big)
+\rho\,\big(\ba\cdot\bx^*-\half\ba^2{t^*}^2\big).
\end{equation}
The extra term, here, is not a total divergence. Accelerations are,
therefore, not symmetries for the fluid equations. In fact,
they carry  the system into an accelerated one \cite{Fouxon,HZ}.

This is consistent with the fact that CGA-type
symmetries require masslessness, while fluid mechanics
has nonzero mass; see (\ref{fluidSchr}).

\medskip\noindent
\textbf{Time-dilations: $z=\infty$}

Recall that the free system has actually an
$\SO(d+1,2)$ \textit{dynamical relativistic conformal symmetry}
 --- see \cite{fluid} ---, which is broken to its Poincar\'e
subgroup in $d+1$ dimensions in the Chaplygin case \cite{JackiwFluid,fluid,Bazeia},
\begin{equation}
V(\rho)=\frac{c}{\rho}.
\label{Chappot}
\end{equation}
The Poincar\'e group in $d+1$ dimensions contains the one-parameter centrally extended Galilei group in $d$ dimensions, augmented with time-dilations
\begin{equation}
t^*=\lambda t,
\qquad
\bx^*=\bx
\label{timedil}
\end{equation}
as a subgroup.\footnote{The transformation (\ref{timedil})
can be viewed as the limiting case, $z\to\infty$,
of $z$-dilation.}
Implemented as
\begin{eqnarray}
\theta^*(t,\bx)&=&\lambda\theta(t^*,\bx^*),
\label{Chapthetaimp}
\\[6pt]
\rho^*(t,\bx)&=&\lambda^{-1}\rho(t^*,\bx^*)
\label{Chaprhoimp}
\end{eqnarray}
they provide a symmetry for the free system: indeed, $L\mapsto\lambda L$ is
compensated  for by the transformation law $dtd\bx=\lambda^{-1}dt^*d\bx^*$.
Space dilations and expansions are broken.

Moreover, the only potential consistent with (\ref{timedil}) is
(\ref{Chappot}), that of the Chaplygin gas
\cite{Bazeia,fluid,JackiwFluid}.

Time dilations (\ref{timedil}) act infinitesimally on the fields according to
\begin{equation}
\delta\rho=-\rho+t\,\p_t\rho,
\qquad
\delta\theta=\theta+t\,\p_t\theta.
\end{equation}
We find
$\delta{}L=\p_t(tL)$, so that
the conserved quantity (\ref{consquant}) associated with (\ref{timedil}),
found by the Noether theorem, is therefore
\begin{equation}
\Delta=tH-\int{\!d\bx\,\rho\,\theta}
\label{timedilcons}
\end{equation}
where $H$ is the energy in (\ref{fluidSchr}),
with $V(\rho)$ as in (\ref{Chappot}).
The conservation of (\ref{timedilcons}) can also
be checked directly, using the equations of motion.

\goodbreak

\subsection{Galilean massless particles}\label{Gal30sSection}

Concerning the second, ``Conformal Galilean (CGA)-type''  symmetries, the situation is more subtle.
The  natural candidates are the massless Galilean systems, studied by Souriau forty years ago \cite{Sou}.
In geometrical optics, a classical ``light ray'' can, in fact, be identified with an \textit{oriented straight line}, $\cD$, in Euclidean space~$\bbR^3$. Such a line is characterized by an arbitrary point $\bx\in\cD$, and its direction, i.e., a unit vector, $\bu$, along $\cD$. The manifold of light rays is readily identified with the (co)tangent bundle~$TS^2$ endowed with its canonical symplectic structure, or a twisted symplectic structure if spin is admitted. This model is based on the Euclidean group.

There exists, indeed, a Galilean version of Euclidean ``spinoptics''. The homogeneous symplectic manifolds of the Galilei group to consider are ``massless'', i.e., defined by the invariants $m=0$, $s\neq0$, and $k>0$, a new Galilei-invariant \cite{Sou,GS}.
 
A natural ``evolution space'' for these massless models is $\cV=\bbR\times\bbR^3\times\bbR\times{}S^2$ described by the quadruples $y=(t,\bx,E,\bu)$
and endowed with the closed two-form
\begin{equation}
\sigma=k\,du_A\wedge dx^A-dE\wedge{}dt
-\frac{s}{2}\epsilon_{ABC}\,u^Adu^B\wedge{}du^C
\label{lightsigma}
\end{equation}
where the constants $k>0$, and $s$, are the \textit{color} and the \textit{spin}, respectively.\footnote{
For $s=0$ we get a ``Fermat particle'', i.e., ``spinless light,'' described by the Fermat principle \cite{Sou,SpinOptics}. 
Quantization requires that $s/\hbar$ be a half-integer,  
and the color becomes $k=2\pi\hbar/\lambda$, where~$\lambda$ is the wavelength~\cite{Sou,GS}.}
 
The motions of these massless particles, e.g., the classical ``light rays'', identified with the characteristic curves of the two-form (\ref{lightsigma}), project onto spacetime as oriented \textit{lightlike} straight worldlines directed along $\bu$, viz.,
\begin{equation}
\dt=0,
\qquad
\dot{\bx}=\bu,
\qquad
\dot{E}=0,
\qquad
\dot{\bu}=0
\label{lighteqmot}
\end{equation}
where we have chosen the parameter $\tau$ as the arc-length 
along the straight line $\cD$. 

\goodbreak

Such a ``motion'' is \textit{instantaneous},
\begin{equation}
t=\const
\label{lightlike}
\end{equation}
and, hence, projects as a lightlike geodesic (\ref{spacelikegeod}) of flat Newton-Cartan spacetime (massless particles have ``infinite speed'').

By the very construction of the model, the Galilei Lie algebra, $\gal(3)$, see (\ref{gal}), acts in a Hamiltonian way (\ref{presympNoether}) on the
evolution space $\cV$ according to
\begin{eqnarray}
\wX=\varepsilon\frac{\partial}{\partial{}t}
+
(\omega^A_B\,x^B+\beta^At+
\gamma^A)\frac{\partial}{\partial{}x^A}
+
ku_A\beta^A\frac{\partial}{\partial{}E}
+
\omega^A_B\,u^B\frac{\partial}{\partial{}u^A},
\label{lightGalac}
\end{eqnarray}
where $\bomega\in\so(3)$, $\bbeta,\bgamma\in\bbR^3$, and $\varepsilon\in\bbR$.

\goodbreak

Using the definition (\ref{Jm}) of the Hamiltonian, $\cJ_\wX$, of this action, we find the associated conserved quantities, namely
\begin{equation}
\begin{array}{llll}
\bP&=&k\bu&\hbox{\small Linear momentum}
\\[8pt]
\bg&=&-\bP t&\hbox{\small Galilean boost}
\\[8pt]
\bJ&=&\bx\times\bP+s\bu\qquad\qquad&\hbox{\small Angular momentum}
\\[8pt]
H&=&E\displaystyle
&\hbox{Energy}
\end{array}
\label{0sGalConsQuant}
\end{equation}

Let us emphasize that the time-dependent Galilei boost, $\bG$,
\textit{is} a constant of the ``motion'', since the  latter takes
place at constant time, cf. (\ref{lightlike}).
Under the Poisson bracket defined by the induced symplectic two-form $\Omega=k\,du_A\wedge{}dq^A-dE\wedge{}dt$, where $\bq=-\bu\times(\bu\times\bx)$, on the space of motions $\cU=TS^2\times{}T\bbR$, the components~(\ref{0sGalConsQuant}) of the moment map close into the \textit{centerless} Galilei group.
In particular, translations and Galilean boosts commute:
our ``photon'' is massless. Curiously, the energy, $E$, remains arbitrary, and determined by the initial conditions.  
 
What about our conformal 
extensions? Are they symmetries? For the \textit{trajectories}, the answer is positive: the lightlike ``instantaneous'' geodesics are permuted by construction, see Section \ref{NCCLightSection}.

 Concerning the \textit{dynamics},
the answer is more subtle though: 
for any finite dynamical exponent $z$, \textit{none} of the
additional geometric symmetries leaves the dynamics invariant. Consider, for example, a dilation: while 
 $\bx\mapsto e^{\lambda}\bx$, the unit vector, $\bu$,
 cannot be dilated,
 $\bu\mapsto\bu$. Therefore,
 a ``photon'' of color $k$ is carried into one with color 
 $ke^{-\lambda}$ (which, in empty space, follows the same trajectories). 
 
There is, however, a way to escape this obstruction:
it is enough \dots \textit{not to dilate}~$\bx$!
To see this, consider
first the spinless ``Fermat'' case. Then the evolution space can be viewed as the submanifold  $\cV\subset{}T^*M$ of the cotangent bundle of spacetime $M=\bbR\times\bbR^3$ defined by the equation
\begin{equation}
\gamma^{ab}p_ap_b-k^2=0.
\label{liftsurface}
\end{equation}
Its presymplectic two-form, given by Equation (\ref{lightsigma}) with $s=0$, is just $\sigma=d\varpi$, where $\varpi=p_adx^a$ is the
restriction to $\cV$ of the canonical one-form of $T^*M$.
Recall that a vector field, $X$, on space-time is \textit{canonically} lifted 
to $T^*M$ as 
\begin{equation}
\wX=X^a\frac{\partial}{\partial{}x^a}-p_b\frac{\partial{X^b}}{\partial{}x^a}\frac{\partial}{\partial{}p_a}.
\label{canlift}
\end{equation}

\goodbreak

One easily sees that this lift is tangent to
the submanifold $\cV$, cf. (\ref{liftsurface}), iff one has identically $\wX(\gamma^{ab}p_ap_b-k^2)=(L_X\gamma)^{ab}p_ap_b=0$, i.e., iff the vector field $X$ leaves~$\gamma$ invariant, viz.,\footnote{This construction is general, and can be extended to the case of any NC-structure \cite{Duv0}.}
\begin{equation}
L_X\gamma=0
\label{m0cond}
\end{equation} 
which is the Galilei-conformal condition
(\ref{galconfMN}) with $m=1$ and $n=0$; the dynamical
exponent is therefore $z=\infty$. 

In the flat case under study, the general solution of Equation (\ref{m0cond}) is given by~(\ref{zinfinity}), i.e.,~by
\begin{equation}
X=\xi(t)\frac{\partial}{\partial{}t}
+\left(\omega^A_B(t)x^B+\eta^A(t)\right)\frac{\partial}{\partial{}x^A}
\label{zinf}
\end{equation}
where $\bomega(t)\in\so(3)$, $\veta(t)$, and $\xi(t)$ depend arbitrarily on time. At last, the maximal Hamiltonian symmetries of the ``Fermat'' particle model constitute the Lie algebra~$\cgal_\infty(3)$, i.e., an infinite-dimensional conformal
extension of the (centerless) Galilei group.
 
Let us compute the explicit form of the canonical lift,~$\wX$, of $X\in\cgal_\infty(3)$ to~$\cV$. Using Equation (\ref{canlift}), we end up with
\begin{eqnarray}
\wX
&=&\xi(t)\frac{\partial}{\partial{}t}
+\left(\omega^A_B(t)x^B+\eta^A(t)\right)\frac{\partial}{\partial{}x^A}\nonumber\\[6pt]
&&+\left(
k(\omega'_{AB}(t)u^Ax^B+\eta'_A(t)u^A)-\xi'(t)E
\right)\frac{\partial}{\partial{}E}\\[6pt]
\nonumber
&&+\omega^A_B(t)u^B\frac{\partial}{\partial{}u^A}
\label{zinflift}
\end{eqnarray}
with the same notation as before. As previously mentioned, the $\cgal_\infty(3)$-action on~$\cV$ is
Hamiltonian if $s=0$. Using Equation (\ref{presympNoether}), one finds the conserved Hamiltonian
\begin{equation}
\cJ_\wX
=(\bx\times{}k\bu)\cdot\bomega(t)+k\bu\cdot\veta(t)-\xi(t)E.
\label{J00}
\end{equation}

What about spin? One easily checks that, in the case $s\neq0$, the presymplectic two-form (\ref{lightsigma}) is no longer $\cgal_\infty(3)$-invariant. In fact, elementary calculation shows that any $X\in\cgal_\infty(3)$ such that $L_\wX\sigma=0$ is of the form (\ref{zinf}) with
\begin{equation}
\bomega'(t)=0.
\label{omegaconst}
\end{equation}

\goodbreak

Thus, in the general case of massless, spinning Galilean particles, the associated constants of the ``motion'' retain the final form
\begin{equation}
\cJ_\wX
=(\bx\times{}k\bu+s\bu)\cdot\bomega+k\bu\cdot\veta(t)-\xi(t)E
\label{J0s}
\end{equation}
with $\bomega\in\so(3)$, $\veta(t)$, and $\xi(t)$ remaining arbitrary functions of time. Note that the conservation of these quantities is related to the fact that the ``motions'' are instantaneous (\ref{lightlike}).

In conclusion, Souriau's ``classical photon'' admits an infinite-dimensional conformal extension of the (centerless) Galilei group; see (\ref{0sGalConsQuant}).

Let us mention, for completeness, another type of massless Galilean particle, introduced by Stichel and Zakrzewski \cite{SZcosmo}. It is
 described by an extended phase space and, unlike Souriau's photon, has \emph{finite} velocity.
It realizes dynamically the Conformal Galilean (CG) symmetry.

\subsection{Galilean Electromagnetism}

Le Bellac and L\'evy-Leblond (LBLL) \cite{LBLL} have discovered, in the early seventies, a full-fledged theory of non-relativistic electromagnetism. They have, actually, highlighted the existence of \textit{two} quite distinct Galilean electromagnetisms, namely a \textit{magnetic-like} and an \textit{electric-like} theory that stem from different non-relativistic limits of Maxwell's theory. The LBLL theories have been, since then, cast into the geometric structure of NC-spacetime~\cite{Kun1}. They have, likewise, been formulated in the ``null Kaluza-Klein'' (or Bargmann) framework of non-relativistic spacetime \cite{DGH}.

Let us, here, confine considerations to the magnetic-like LBLL theory along the lines of \cite{Kun1}. Given a (d+1)-dimensional NC-spacetime structure $(M,\gamma,\theta,\Gamma)$, it is described by the following couple of PDE, namely
\begin{eqnarray}
\label{dF=0}
d\sF&=&0\\
\label{divF=J}
\rdiv\sF&=&J
\end{eqnarray} 
involving a two-form, $\sF=\half \sF_{ab}\,dx^a\wedge{}dx^b$, of $M$ interpreted as the 
electromagnetic field, and a one-form, $J$, the current density of the sources. In Equation~(\ref{divF=J}), one must read
\begin{equation}
\rdiv\sF_c=\gamma^{ab}\nabla_a\sF_{bc}
\label{defdiv}
\end{equation}
for all $c=0,\ldots,d$. 

Note that $d=3$ in the original formulation of LBLL theory where equations (\ref{dF=0}), and (\ref{divF=J}) retain the form
$$
\bnabla\cdot\bB=0,
\qquad
\bnabla\times\bE+\frac{\partial\bB}{\partial t}=0
$$
and 
$$
\bnabla\cdot\bE=\varrho,
\qquad
\bnabla\times\bB=\bj
$$
respectively, once we posit $E_A=\cF_{A0}$, and $B^A=\half\epsilon^{ABC}\cF_{BC}$, for the components of the electromagnetic field, as well as $\varrho=J_0$, and $j_A=J_A$ for the those of the current density, with $A=1,2,3$. 

\goodbreak

Notice the absence
of  the displacement current in Amp\`ere's law:
its presence would, clearly, break the Galilean
symmetry (Maxwell's equations are relativistic).


Let us show that, much in the same way as Maxwell's sourcefree electro\-magnetism, the maximal symmetries of the sourcefree LBLL magnetic theory are actually richer than those expected from the original spacetime structure. More specifically, let us look at all conformal Galilei transformations that preserve the LBLL Equations (\ref{dF=0}) and (\ref{divF=J}), with $J=0$. We will thus seek the maximal Lie algebra of Galilei conformal vector fields $X$ of $(M,\gamma,\theta)$, i.e., satisfying (\ref{confgal}), and such that
\begin{eqnarray}
\label{LXdF=dLXF}
L_Xd\sF&=&dL_X\sF\\
\label{LXdivF=divLXF}
L_X\rdiv\sF&=&\rdiv{}L_X\sF
\end{eqnarray} 
for all solutions, $\sF$, of the above sourcefree LBLL equations.

Equation (\ref{LXdF=dLXF}) is trivially satisfied (as a consequence of the general fact that the Lie and exterior derivatives commute). As to Equation (\ref{LXdivF=divLXF}), one finds
\begin{eqnarray}
0
\nonumber
&=&
L_X\gamma^{ab}\nabla_a\sF_{bc}-\gamma^{ab}\nabla_aL_X\sF_{bc}\\
\nonumber
&=&
(L_X\gamma)^{ab}\nabla_a\sF_{bc}-\gamma^{ab}((L_X\Gamma)^k_{ab}\sF^{}_{kc}+(L_X\Gamma)^k_{ac}\sF^{}_{bk})\\
\nonumber
&=&
f\gamma^{ab}\nabla_a\sF_{bc}+2\gamma^{ab}(L_X\Gamma)^k_{a[b}\sF^{}_{c]k}\\
&=&
-2\gamma^{ab}(L_X\Gamma)^k_{a[b}\delta^\ell_{c]}\sF^{}_{k\ell}
\end{eqnarray}
since $L_X\gamma^{ab}=f\gamma^{ab}$, and $\gamma^{ab}\nabla_a\sF_{bc}=0$. This readily entails 
\begin{equation}
\gamma^{ab}(L_X\Gamma)^{[k}_{a[b}\delta^{\ell]}_{c]}=0
\label{LBLLInvCond}
\end{equation}
for all $c,k,\ell=0,\ldots,d$.

\goodbreak

Utilizing Equation (\ref{deltaGamma}), giving  the most general form of the variations of the NC-connection compatible with Galilei conformal rescalings, we will now put $\delta\Gamma^c_{ab}=L_X\Gamma^c_{ab}$, and easily show that Equation (\ref{LBLLInvCond}) writes now
\begin{equation}
\left((4-d)\delta^{[k}_c
-2\theta_cU^{[k}\right)\gamma^{\ell]a}\partial_a f
+(f+g)\theta_cF_{ab}\gamma^{ka}\gamma^{\ell{}b}=0.
\label{LBLLInvCondBis}
\end{equation}
Taking traces, we readily deduce that $\gamma^{\ell{}a}\partial_af=0$, hence that Equation (\ref{LBLLInvCondBis}) reads
\begin{equation}
(f+g)F_{ab}\gamma^{ka}\gamma^{\ell{}b}=0.
\label{LBLLInvCondTer}
\end{equation}

On the one hand, we can have the case $f+g=0$ (with $z=2$) leading us to the Schr\"odinger Lie algebra, $\sch(M,\gamma,\theta,\Gamma)$, for a general NC-structure.

On the other hand, considering a NC-Milne structure $(M,\gamma,\theta,\GammaU)$, charac\-terized by $F=0$, already enables us to satisfy Equation (\ref{LBLLInvCondTer}), hence the full system (\ref{LXgammaBis})--(\ref{LXGammaBis}) providing us with a higher-dimensional symmetry algebra.

\goodbreak

We have just proved that the maximal Lie algebra of symmetries of the Le Bellac-L\'evy-Leblond equations in vacuum is isomorphic to $\cmil(M,\gamma,\theta,\GammaU)$. In the flat case this is the Lie algebra $\cmil(d)$ --- see (\ref{cmild}) ---, containing the CGA  (\ref{CGA}).

\section{Conclusion}

In this paper, we have presented a systematic way to derive
all types of ``non-relativistic conformal transformations'' of spacetime $M$. Due
to the degeneracy of the Galilei ``metric'' $(\gamma,\theta)$, and to the relative independence of Newton-Cartan connections, $\Gamma$, there are quite a large number of candidates.

Firstly, the conformal transformations of the ``metric''
structure alone, (\ref{confgal}), yield the conformal Galilei Lie 
algebra $\cgal(M,\gamma,\theta)$. In the flat case, and in $d$ 
space dimensions, it is the infinite-dimensional Lie algebra
 (\ref{cgal}). Fixing the dynamical exponent, $z$, via the geometric definition (\ref{galconfMN}), yields a family of (still infinite-dimensional) Lie subalgebras,
 $\cgal_z(d)$. For $z=2$, we get the Schr\"odinger-Virasoro Lie
 algebra~(\ref{sv}) of Henkel et al. \cite{Hen,Henkel03,Henkel06}, and for $z=\infty$ we get $\cgal_\infty(d)$ in~(\ref{zinfinity}).
 
Secondly, the Newton-Cartan structure also involves the choice of a connection,
which allows us to consider the symmetries of the equations of geodesics, identified with worldlines of test particles. 

- The conformal Galilei transformations
which  permute geodesics, which are generically \textit{timelike}, are,
in fact, Schr\"odinger transformations. Those
with dynamical exponent $z=2$ constitute the Schr\"odinger Lie algebra, $\sch(d)$, see (\ref{sch2d}), in the case of flat NC-spacetime. 

- Those which exchange \textit{lightlike} geodesics have a
richer structure, though.  The resulting infinite-dimensional algebra, $\cnc(d)$, is given by (\ref{cncd}) in the flat case. There is no \textit{a priori}
restriction on the dynamical exponent, $z$.
The infinite-dimensional Lie algebra~$\cnc(d)$ admits a finite-dimensional Lie subalgebra, $\cmil(d)$ --- related to a flat \textit{NC-Milne structure} --- featuring independent space and time dilations, as well as new ``acceleration'' generators; see (\ref{cmild}).
The Lie subalgebras associated with a dynamical exponent, $z$, are respectively (i) the Schr\"odinger algebra, for $z=2$, and (ii) the CGA (\ref{CGA}) of Henkel \cite{Hen} and Lukierski et al. \cite{LSZGalconf}, for~$z=1$. The infinite-dimensional Lie algebra $\cnc_\infty(d)$ completes our classification.

Our geometric-algebraic framework, dealing with vector fields on NC-space\-time, leaves
no place to central extensions; the latter only arise when
conserved quantities --- or (pre)quantization --- of concrete
physical systems are considered.

All these symmetries were derived by considering as fundamental the NC-structure of non-relativistic spacetime. What about concrete physical systems? In Section \ref{dynSys} we study two such systems. Souriau's (pre)symplectic framework \cite{Sou} allows us to present them in a unified manner. 

\goodbreak
 
The dynamics of physical systems reduce further
the geometric symmetries to some of their subgroups. This is
understood if we think of free fall: massive
particles fall in the same way, independently of their respective
masses. The trajectory of a particle can be carried therefore 
into another one by
a geometric transformation. However, such a transformation can
change the mass --- so it is \textit{not} necessarily a symmetry of
the dynamical system.

Firstly, for a massive Galilean particle with spin, we recover the well-known Schr\"odinger symmetry associated with $\sch(3)$.

Secondly, the hydrodynamics of irrotational fluids turns out to provide a special instance of a classical field theory invariant under the Schr\"odinger Lie algebra, $\sch(d)$, yielding new conserved quantities, apart from the standard Galilean constants of the motion.

As to the GCA-type symmetries, they require to have no mass \cite{LSZGalconf}. The natural candidates are, therefore, Souriau's ``Galilean photons'' \cite{Sou}, which
are associated with the coadjoint orbits of the (centerless)
Galilei group. These ``particles'' have an  ``instanta\-neous motion'' --- they have ``infinite velocity''. They can
carry spin, generalizing ``spinless light'',  described
by the Fermat principle \cite{Sou,SpinOptics}.

Can one take such models seriously? The answer is positive since they can be
obtained as suitable non-relativistic limits of relativistic
massless particles, i.e., those associated with the
mass zero coadjoint orbits of the Poincar\'e group \cite{Sou}. Even more importantly, the model of ``Galilean photon'' ($s=\pm\hbar$)
is a trivial extension of the Euclidean model presented in \cite{SpinOptics}, which has been used to explain the recently observed spin-Hall effect for light \cite{OHE}.

Let us emphasize  that Souriau's model of massless 
Galilean particles carry an infinite-dimensional Lie algebra of symmetries, namely $\cgal_{\infty}(3)$. 
\goodbreak

As a provisionally last illustration of our formalism, we show that the maximal symmetry Lie algebra of the Le Bellac-L\'evy-Leblond equations of (magnetic-like) Galilean electromagnetism in vacuum turns out to be the conformal NC-Milne algebra, i.e., $\cmil(d)$, in flat spacetime, with the CGA as a Lie subalgebra.

Let us also refer to \cite{RU2} where the space of periodic time-dependent Schr\"odinger operators, in the case $d=1$, has been shown to be naturally $\sv(1)$-invariant.

Recently, a supersymmetric extension of the CGA has been found \cite{AzLuk}.

Let us end our paper with some historical remarks, cf. \cite{Hoss}.

The first person to consider the CGA seems to be
Barut \cite{Barut}, in 1973, who derived it by contraction from the relativistic conformal group.
But then he discarded it, however,
arguing that it is not a symmetry of the Schr\"odinger equation.
In 1978,  Havas and Pleba\'nski generalized both the Schr\"odinger and CG groups
to an infinite-dimensional group \cite{HaPl}.

Even more astonishingly, the Schr\"odinger symmetry has
already be known to Jacobi  \cite{HaPl}. 
In his 1842/43 lectures delivered at the University of K\"onigsberg \cite{Jacobi},
he studied indeed the dynamics of a particle in a homogeneous
potential, $U$, of degree~$k$. Using a  scaling argument reminiscent of the proof of the virial theorem (see, e.g., \cite{DGH}), he proved
that
\begin{equation}
\frac{\ d^2}{dt^2}\Big(\frac{m\bx^2}{2}\Big)=-(k+2)U+2E
\label{Jacrel1}
\end{equation}
where
$E$ is, in modern terms, the conserved  energy. Then he observed that
for the \emph{inverse-square potential}, $k=-2$, Equation (\ref{Jacrel1}) can be rewritten as
$$
\frac{\ d}{dt}\left(m\bx\cdot\dot{\bx}-2Et\right)=0.
$$
Putting $\bp=m\dot{\bx}$, the quantity
\begin{equation}
D=\bp\cdot\bx-2Et
\label{Jacdil}
\end{equation}
is therefore conserved. Then 
$$
\frac{\ d}{dt}\left(\frac{m\bx^2}{2}-tD-Et^2\right)=0
$$
so that
\begin{equation}
K=\frac{m\bx^2}{2}-tD-Et^2
\label{Jacexp}
\end{equation}
is also conserved. But $E,\,D,\,K$ are precisely the conserved
quantities which stem, through the Noether theorem, from the conformal,
$\Ort(2,1)$, subgroup of the Schr\"o\-dinger group, cf. (\ref{msConsQuant}). 
 
\parag
\textbf{Acknowledgement.}\
We are greatly indebted to M. Henkel, J. Lukierski, P. Stichel, and S. Rouhani for useful cor\-respondence. 





\end{document}